\documentclass[12pt]{article}   
\usepackage{amsmath,amsthm,latexsym,amssymb,amsfonts}
\allowdisplaybreaks                                  
\usepackage{epsfig}
\makeatletter
\@addtoreset{equation}{section}
\makeatother

\newtheorem{Theorem}{Theorem}[section]
\newtheorem{Definition}{Definition}[section]
\newtheorem{Lemma}{Lemma}[section]

\newcommand{\evec}[1]{\vec{#1}}
\newcommand{\betr}[1]{\left\lvert #1 \right\rvert}    
\newcommand{\cc}[1]{\overline{#1}} 
\newcommand{\scpr}[2]{\langle#1,#2\rangle} 
\newcommand{\comm}[2]{\left[#1,#2\right]} 
\newcommand{\ket}[1]{\lvert #1\rangle} 

\newcommand{\lzwo}{L^2} 
\newcommand{\aver}[1]{\langle\langle#1\rangle\rangle}  
\newcommand{\expec}[1]{\langle#1\rangle}

\newcommand{\qqquad}{\qquad\qquad}

\newcommand{\hilb}[1]{\mathcal{#1}}
\newcommand{\teil}[1]{\vspace{1ex plus0.5ex minus0.5ex}
{\bf #1}\\[0.6ex plus0.3ex minus0.3ex]}

\newcommand{\vla}[0]{\xrightarrow{\hspace{5ex}}}
\newcommand{\R}{\mathbb{R}}              
\newcommand{\Z}{\mathbb{Z}}              
\newcommand{\C}{\mathbb{C}}              
\newcommand{\N}{\mathbb{N}}              

\newcommand{\I}{\text{I}}

\newcommand{\one}{1}

\newcommand{\con}{\mathcal{A}}   
\newcommand{\gcon}{\overline{\mathcal{A}}} 
\DeclareMathSymbol{\comp}{\mathrel}{AMSa}{"62} 
\DeclareMathOperator{\tr}{Tr}

\DeclareMathOperator{\cyl}{Cyl}
\DeclareMathOperator{\E}{\mathcal{E}}
\DeclareMathOperator{\porder}{\mathcal{P}}
\DeclareMathOperator{\spa}{span}
\def\be#1\ee{\begin{equation}#1\end{equation}}
\def\ba#1\ea{\begin{eqnarray}#1\end{eqnarray}}
\def\a{{\cal A}}
\def\ab{\overline{\a}}

\def\Nl{{\mathchoice
{\setbox0=\hbox{$\displaystyle\rm N$}\hbox{\hbox to0pt
{\kern0.4\wd0\vrule height0.9\ht0\hss}\box0}}
{\setbox0=\hbox{$\textstyle\rm N$}\hbox{\hbox to0pt
{\kern0.4\wd0\vrule height0.9\ht0\hss}\box0}}
{\setbox0=\hbox{$\scriptstyle\rm N$}\hbox{\hbox to0pt
{\kern0.4\wd0\vrule height0.9\ht0\hss}\box0}}
{\setbox0=\hbox{$\scriptscriptstyle\rm N$}\hbox{\hbox to0pt
{\kern0.4\wd0\vrule height0.9\ht0\hss}\box0}}}}
\def\Zl{{\mathchoice
{\setbox0=\hbox{$\displaystyle\rm Z$}\hbox{\hbox to0pt
{\kern0.4\wd0\vrule height0.9\ht0\hss}\box0}}
{\setbox0=\hbox{$\textstyle\rm Z$}\hbox{\hbox to0pt
{\kern0.4\wd0\vrule height0.9\ht0\hss}\box0}}
{\setbox0=\hbox{$\scriptstyle\rm Z$}\hbox{\hbox to0pt
{\kern0.4\wd0\vrule height0.9\ht0\hss}\box0}}
{\setbox0=\hbox{$\scriptscriptstyle\rm Z$}\hbox{\hbox to0pt
{\kern0.4\wd0\vrule height0.9\ht0\hss}\box0}}}}
\def\Ql{{\mathchoice
{\setbox0=\hbox{$\displaystyle\rm Q$}\hbox{\hbox to0pt
{\kern0.4\wd0\vrule height0.9\ht0\hss}\box0}}
{\setbox0=\hbox{$\textstyle\rm Q$}\hbox{\hbox to0pt
{\kern0.4\wd0\vrule height0.9\ht0\hss}\box0}}
{\setbox0=\hbox{$\scriptstyle\rm Q$}\hbox{\hbox to0pt
{\kern0.4\wd0\vrule height0.9\ht0\hss}\box0}}
{\setbox0=\hbox{$\scriptscriptstyle\rm Q$}\hbox{\hbox to0pt
{\kern0.4\wd0\vrule height0.9\ht0\hss}\box0}}}}
\def\Rl{{\mathchoice
{\setbox0=\hbox{$\displaystyle\rm R$}\hbox{\hbox to0pt
{\kern0.4\wd0\vrule height0.9\ht0\hss}\box0}}
{\setbox0=\hbox{$\textstyle\rm R$}\hbox{\hbox to0pt
{\kern0.4\wd0\vrule height0.9\ht0\hss}\box0}}
{\setbox0=\hbox{$\scriptstyle\rm R$}\hbox{\hbox to0pt
{\kern0.4\wd0\vrule height0.9\ht0\hss}\box0}}
{\setbox0=\hbox{$\scriptscriptstyle\rm R$}\hbox{\hbox to0pt
{\kern0.4\wd0\vrule height0.9\ht0\hss}\box0}}}}
\def\Cl{{\mathchoice
{\setbox0=\hbox{$\displaystyle\rm C$}\hbox{\hbox to0pt
{\kern0.4\wd0\vrule height0.9\ht0\hss}\box0}}
{\setbox0=\hbox{$\textstyle\rm C$}\hbox{\hbox to0pt
{\kern0.4\wd0\vrule height0.9\ht0\hss}\box0}}
{\setbox0=\hbox{$\scriptstyle\rm C$}\hbox{\hbox to0pt
{\kern0.4\wd0\vrule height0.9\ht0\hss}\box0}}
{\setbox0=\hbox{$\scriptscriptstyle\rm C$}\hbox{\hbox to0pt
{\kern0.4\wd0\vrule height0.9\ht0\hss}\box0}}}}
\def\Hl{{\mathchoice
{\setbox0=\hbox{$\displaystyle\rm H$}\hbox{\hbox to0pt
{\kern0.4\wd0\vrule height0.9\ht0\hss}\box0}}
{\setbox0=\hbox{$\textstyle\rm H$}\hbox{\hbox to0pt
{\kern0.4\wd0\vrule height0.9\ht0\hss}\box0}}
{\setbox0=\hbox{$\scriptstyle\rm H$}\hbox{\hbox to0pt
{\kern0.4\wd0\vrule height0.9\ht0\hss}\box0}}
{\setbox0=\hbox{$\scriptscriptstyle\rm H$}\hbox{\hbox to0pt
{\kern0.4\wd0\vrule height0.9\ht0\hss}\box0}}}}
\def\Ol{{\mathchoice
{\setbox0=\hbox{$\displaystyle\rm O$}\hbox{\hbox to0pt
{\kern0.4\wd0\vrule height0.9\ht0\hss}\box0}}
{\setbox0=\hbox{$\textstyle\rm O$}\hbox{\hbox to0pt
{\kern0.4\wd0\vrule height0.9\ht0\hss}\box0}}
{\setbox0=\hbox{$\scriptstyle\rm O$}\hbox{\hbox to0pt
{\kern0.4\wd0\vrule height0.9\ht0\hss}\box0}}
{\setbox0=\hbox{$\scriptscriptstyle\rm O$}\hbox{\hbox to0pt
{\kern0.4\wd0\vrule height0.9\ht0\hss}\box0}}}}
\begin{document}
\title{Towards the QFT on Curved Spacetime Limit of QGR.\\I: A General Scheme}
\author{
H. Sahlmann\thanks{sahlmann@aei-potsdam.mpg.de}, 
T. Thiemann\thanks{thiemann@aei-potsdam.mpg.de} \\
       MPI f. Gravitationsphysik, Albert-Einstein-Institut, \\
           Am M\"uhlenberg 1, 14476 Golm near Potsdam, Germany}
\date{{\small PACS No. 04.60, Preprint AEI-2002-049}}

\maketitle

\begin{abstract}
In this article and the companion paper \cite{ST02} we address 
the question of how one might obtain the semiclassical limit of ordinary
matter quantum fields (QFT) propagating on curved spacetimes (CST) from
full fledged Quantum General Relativity (QGR), starting from first 
principles. We stress that we do not claim to have a satisfactory answer
to this question, rather our intention is to ignite a discussion by 
displaying the problems that have to be solved when carrying out such a 
program. 

In the first paper of this series of two we propose a general 
scheme of logical steps that one has to take in order to arrive at such 
a limit. We discuss the technical and conceptual problems that
arise in doing so and how they can be solved in principle. As to be 
expected, completely new issues arise due to the fact that QGR is a 
background independent theory. For instance, fundamentally the notion of 
a photon involves not only the Maxwell quantum field but also the metric
operator -- in a sense, there is no photon vacuum state but a ``photon   
vacuum operator''! Such problems have, to the best of our knowledge, not
been discussed in the literature before, we are facing squarely one 
aspect of the deep 
conceptual difference between a background dependent and a background free 
theory.

While in this first paper we focus on conceptual and abstract aspects, 
for instance the definition of (fundamental) $n-$particle states (e.g. 
photons), in the second paper 
we perform detailed calculations including, among other things, coherent 
state expectation values and propagation on random lattices.  
These calculations serve as an 
illustration of how far one can get with present mathematical techniques.
Although they result in detailed predictions for the size of first quantum 
corrections such as the $\gamma$-ray burst effect, these 
predictions should not be taken too seriously because a) the 
calculations are carried out at the kinematical level only and b)
while we can classify the amount of freedom 
in our constructions, the analysis of the physical significance 
of possible choices has just begun.      
\end{abstract}
\section{Introduction}
\label{se1}
Canonical, non-perturbative Quantum General Relativity (QGR) has by now
reached the status of a serious candidate for a quantum theory of the 
gravitational field:
First of all, the formulation of the theory is mathematically rigorous. 
Although there are 
no further inputs other than the fundamental principles of four-dimensional, 
Lorentzian General Relativity and quantum theory, the  
theory predicts that there is a built in {\it fundamental discreteness} 
at Planck scale distances and therefore an UV cut-off precisely due to its
diffeomorphism invariance (background independence). Next, while most of the 
results have so far been obtained using the canonical operator language,
also a path integral formulation (``spin foams'') is currently 
constructed. Furthermore, as a first physical application, a rigorous, 
microscopical derivation of the Bekenstein-Hawking entropy -- area law has 
been established. 
The reader interested in all the technical details of QGR and its 
present status is 
referred to the exhaustive review article \cite{Thiemann:2001yy} and 
references therein, and to \cite{Rovelli:1998yv} for a less technical 
overview. For a comparison with other
approaches to quantum gravity see
\cite{Horowitz:2000sh,Rovelli:1997qj,Rovelli:1999hz}.\\ 

A topic that has recently attracted much attention is to explore the
regime of QGR where the quantized gravitational field behaves ``almost
classical'', i.e. approximately like a given classical solution to the 
field equations. Only if such a regime exists, one can really claim
that QGR is a viable candidate theory for quantum gravity.  
Consequently, efforts have been made to identify so called 
\textit{semiclassical states} in the Hilbert space of QGR, states 
that reproduce a given classical geometry in terms of their
expectation values and in which the quantum mechanical fluctuations are 
small \cite{Ashtekar:1992tm,Bombelli:2000ua,Thiemann:2000bw,Thiemann:2000ca,Thiemann:2000bx}.
Also, it has been investigated how gravitons emerge as carriers of the 
gravitational interaction in the semiclassical regime of the theory
\cite{IR1,IR2,Ashtekar:1991mz}. 
The recent investigation of Madhavan and others \cite{V1,V2,V3,Ashtekar:2001xp} on the
relation between the Fock representations used in conventional quantum 
field theories and the one in QGR further illuminate the relation
between QGR and a perturbative treatment based on gravitons. 

In this and the companion paper \cite{ST02} we would also like to
contribute to the understanding of the semiclassical limit of QGR: 
We will investigate how the theory of quantum matter fields
propagating in a fixed classical background geometry (QFT on CST)
arises as an approximation to the full theory of QGR coupled to
(quantum) matter fields. 
We will show in section \ref{se4} of the present work, 
how, upon choosing a semiclassical state, an effective QFT for the
matter fields can be obtained from a more fundamental theory of QGR
coupled to matter. 
This effective theory turns out to be very similar to standard 
QFT on CST, but still carries an imprint of the discreteness of the 
geometry in QGR as well as of the quantum fluctuations in the 
gravitational field. 

Validating the semiclassical limit of matter coupled to QGR
is not the only motivation for the present work. Since QGR is a
background independent theory, the consistent coupling of matter fields 
requires a quantum field theoretical description of these fields that
differs considerably from that used in ordinary QFT. Therefore, 
another aim of the
present work is to gain some insights into what these differences are
and how matter QFT can be formulated in a setting where also the
gravitational field is quantized. As a main result of the present
paper we show how a theory of matter coupled to quantum geometry can be
formulated within the framework of QGR. Within this theory we identify 
states that can roughly be compared to the $n$-particle states
occurring in ordinary QFT. Their structure is however fundamentally
different as compared to that of the ordinary Fock states: Their
definition also involves operators of the gravitational sector of the
theory!

Similar considerations may be applied to understand the emergence of
gravitons in the semiclassical limit of QGR. The situation there is however 
a bit more complicated since it requires the separation of the
gravitational field in a background- and a graviton-part. 
We refer the reader to \cite{QFTQGR} where a detailed consideration
will be given. 

Finally, due to a better understanding of the phenomenology of quantum 
gravity and the experiments that could lead to its detection (see
\cite{Amelino-Camelia:2002vw} for a recent review) it is an intriguing 
question whether it might already be possible to make predictions for
observable quantum gravity effects based on QGR.  
In order to do so, one has to consider a coupling of the gravitational field 
to matter -- one can not measure the gravitational field directly but
only through its action on other fields. 
Indeed, ground-breaking work on the phenomenology of QGR has been done 
\cite{Gambini:1998it,Alfaro:1999wd,Alfaro:2001rb,Alfaro:2001gk}.  
In these works, corrections to the standard dispersion relations 
for matter fields due to QGR have been obtained. Since we are dealing 
with a theory for matter coupled to QGR in the present work, it is an 
important question whether the results of
\cite{Gambini:1998it,Alfaro:1999wd,Alfaro:2001rb,Alfaro:2001gk} can be 
confirmed in the present setting. We will
discuss the general aspects of this question in section \ref{se5}. In the
companion paper \cite{ST02} we will carry out a more detailed
calculation, based on the results of the present work and the
semiclassical states constructed in
\cite{Thiemann:2000bw,Thiemann:2000ca,Thiemann:2000bx}.\\ 

The main difficulty in carrying out the program outlined up to now lies in 
the fact that the full dynamics of quantum gravity coupled to quantum
matter is highly complicated. This would already be the case for
ordinary interacting fields but is amplified in the present case  
due to the complicated interaction terms (the gravitational field
enters in a non-polynomial way)
and the difficulties in the interpretation of the resulting solutions.
In the setting of QGR, the dynamics is implemented in the spirit of
Dirac, by turning the Hamilton constraint of the classical theory 
into an operator
and restricting attention to (generalized) states in its kernel.  
A mathematically well-defined candidate Hamiltonian constraint operator
has been proposed in \cite{Thiemann:1998rq,Thiemann:1998rv,Thiemann:1998rt,
Thiemann:1998aw,Thiemann:1998av} (see also
\cite{DiBartolo:1999fq,DiBartolo:1999fr} 
for another proposal based on Vasiliev invariants). This operator
turns out to be very complicated and a systematic analysis of its
kernel seems presently out of reach. 
Therefore, in our considerations, we can not start from a fully
quantized dynamical theory of gravity coupled to matter. Instead,  
we have to treat the dynamics in some rather crude 
approximation and therefore {\bf our considerations will be 
kinematical to a large extent.} 
To be more precise, we will \textit{not} treat the the matter parts in 
the Hamiltonian as constraints, but as Hamiltonians generating the 
dynamics of the
matter fields in the ordinary QFT sense. 
With the part in the Hamiltonian describing the self interaction of the
gravitational field we will deal by using semiclassical
states, which, as we will explain, annihilate this part of the
Hamiltonian constraint at least approximately.   
Proceeding in this way certainly only amounts to establishing an
approximation to the full theory: 
The self interaction of gravity and the back-reaction of the matter fields on
the geometry are only partly reflected by using semiclassical states
that approximate a \textit{classical} solution to the field equations
of the gravity-matter system. 

What we gain is a relatively easy to interprete, \textit{fully
  quantized} theory of gravity and matter fields. 
This way we have 
``a foot in the door'' to the fascinating topic of interaction between 
quantum matter and quantum gravity and can start to discuss the
conceptual issues arising, as well as take some steps towards the
prediction of observable effects resulting from this
interplay.\\

Let us finish this introduction with a brief description of the
structure of the rest of the article:
In the next section, we will discuss the main steps taken in this work 
in more detail before we turn to their technical implementation in the
subsequent sections.
 
In section \ref{se3} we give a very brief introduction to the
formalism of QGR, mainly to fix our notation. In more detail we display
the matter Hamiltonian operators of electromagnetic, scalar and Dirac
matter when coupled to general relativity.

Section \ref{se4} contains the main results of this paper, namely a 
proposal for how to
arrive at the notion of Fock states or $n-$particle states on fluctuating 
quantum spacetimes, if one is to start from a fundamental quantum theory of 
gravity of matter.

In section \ref{se5} we discuss various methods to obtain dispersion
relations for the matter fields from the full theory described in
\ref{se3}.

We conclude this work with a discussion of its results and possible
directions for future research in section \ref{se6}.

In an appendix, we treat the toy model of two coupled Harmonic
oscillators to give an example of how the results are affected when
one uses kinematical coherent states instead of coherent states in the 
dynamical Hilbert space of the theory.\\ 

As already said, in the present paper we focus on describing the general
scheme, detailed calculations will appear in the companion 
paper \cite{ST02}.
\section{A General Scheme}
\label{se2}
In this section we want to discuss the issues related to the
QFT on CST Limit of QGR and describe the steps taken in the present
work in more detail.\\ 

The first step that we will take is the kinematical quantization of
the matter and the gravitational field on a Hilbert space
$\hilb{H}^{\text{kin}}$. We will be guided by the 
fundamental principles of QGR which have to be obeyed: The quantum
theory should be formulated in a background free and diffeomorphism
covariant fashion. 
If the matter field is a gauge field with compact
gauge group, we can quantize it with exactly the same methods that are
used in QGR for the gravitational field. This way, we obtain a neat
unified description of gravity and the other gauge fields.  
Also for fermions or 
scalar fields, a representation should be used that is background
independent. This rules out the usual Fock representation.
New representations for fermionic, Higgs and scalar fields in keeping with 
the principles of QGR were proposed in \cite{Thiemann:1998rq} and we 
will use them for our purpose. 

The quantization of the Hamiltonian of the coupled system is
a rather nontrivial task, due to its complicated non-polynomial
dependence on the basic variables of the theory. 
Nevertheless, a scheme for the quantization for densities of weight one
has been proposed in
\cite{Thiemann:1998aw,Thiemann:1998rt} which leads to well-defined 
candidate operators. 
The resulting operators are quite complicated but perfectly well
defined and lead to reasonable results in a symmetry reduced context 
\cite{Bojowald:2002gz,Bojowald:2001vw}. 
Another very encouraging aspect of the scheme is that it works
precisely \textit{due} to the density one character of the classical
quantities, which is dictated by background independence, and not only 
\textit{despite} of it.
In \cite{ST02} and the present paper we will proceed along the lines given in 
\cite{Thiemann:1998aw,Thiemann:1998rt} and obtain Hamiltonian
constraint operators for electromagnetic, scalar, and fermionic fields 
coupled to gravity.\\ 

As a next step we have to deal with the constraints of the theory: 
A Gau{\ss} constraint for Gravity and for every matter gauge field, 
the spatial diffeomorphism constraint of gravity, and, 
finally and most importantly, the Hamilton constraint of the coupled 
gravity-matter system.  

The implementation of the diffeomorphism constraint has been
accomplished in \cite{Ashtekar:1995zh}. 
Still, there is a difficulty related to the spatial diffeomorphism
constraint for \textit{pure} gravity: No spatially diffeomorphism 
invariant quantum observables (apart from the total volume of the 
space-like hypersurface $\Sigma$, in case 
it is finite) have been constructed so 
far. This is due to the fact that such observables are given by 
integrals over $\Sigma$ of scalar densities of weight one built from 
the spatial curvature tensor and its spatial covariant derivatives which
are highly non-polynomial functions.
This problem gets alleviated when matter is coupled to the
gravitational field. 
For instance, the matter can serve to define 
submanifolds or regions of $\Sigma$ in a diffeomorphism covariant
way. Diffeomorphism invariant observables can then be obtained by
integrating the gravitational fields over these submanifolds or regions 
\cite{Rovelli:1991ph,Rovelli:1991pi}.
Indeed we will see that this also applies to the Hamiltonian for
gravity coupled matter: The corresponding operator constructed in the
next section will be diffeomorphism invariant. This is important for
the following reason: 
Since the diffeomorphisms of $\Sigma$ are implemented unitarily
on $\hilb{H}_{\text{kin}}$, the expectation value and fluctuations 
of a diffeomorphism invariant operator do not differ from 
its expectation value and fluctuations in the state that results from 
projecting the
original one to the diffeomorphism invariant Hilbert space (via group
averaging) \cite{Ashtekar:1995zh} provided the operator 
satisfies certain technical conditions (it has to leave cylindrical
subspaces of the Hilbert space separately invariant).
Therefore as long as we work with diffeomorphism invariant 
operators on $\hilb{H}_{\text{kin}}$ we do not have to bother too 
much about implementing the diffeomorphism constraint. 
Similar remarks concern the Gau{\ss} constraints, so we will also not
be concerned with their implementation in what follows.  

We now turn to the implementation of the Hamilton constraint. 
Even for pure gravity, this is a very difficult topic.  
Though solutions have been
found \cite{Thiemann:1998av,Thiemann:1998rv}, they are notoriously
hard to interpret due to the lack of Dirac
observables invariant under the motions generated by the Hamiltonian
constraint (even in the presence of matter) and a
thorough understanding of the ``problem of time''.  The problem of
finding solutions to the Hamilton constraint for gravity coupled to
matter has not been treated before although the method of 
\cite{Thiemann:1998av} can in principle be applied as well.
\footnote{Notice that, since presently the correctness of the classical 
limit of the operators corresponding to the quantization 
of the geometry and matter Hamiltonian constraints 
proposed in \cite{Thiemann:1998aw,Thiemann:1998rt} is not yet confirmed,
in order to verify this proposal 
it is well motivated to work with kinematical semiclassical states:
This is because 
one cannot study the semiclassical limit of an operator on its kernel.
Also, since the spatially diffeomorphism invariant states are not left
invariant by the Hamiltonian constraint, we cannot even work at the 
spatially diffeomorphism invariant level. In this paper we are, however,
not so much interested in testing the Hamiltonian constraint but rather
we suppose that some correct version of it exists and ask how physical
predictions can be extracted without solving the complicated theory 
exactly.}      

Since one of our goals is
to explore the semiclassical limit of QGR coupled to matter, 
the task presented to us is even harder:
Not only do we have to find \textit{some} solutions to the Hamilton
constraint, but we are interested in \textit{specific} solutions in
which the gravitational field is in a state close to some given
classical geometry. 

As already explained in the introduction, in the light of these
difficulties, we propose to proceed along slightly different lines. 
To give an idea what we are
aiming at, imagine we ought to compute corrections to the interaction
of some quantum system (an atom, say) with an electromagnetic field, which 
are due to the quantum nature of the electromagnetic field.
Ultimately this is a problem in quantum
electrodynamics and therefore certainly not solvable in full generality. What
can be done?
For the free Maxwell field, there is a family of
states describing configurations of the quantum field close to
classical ones, the \textit{coherent states}: Expectation values for
field operators yield the classical values and the quantum mechanical
uncertainties are minimal in a specific sense. Such states could be
used to model the classical electromagnetic field.
Certainly these coherent states are
no viable states for the full quantum electrodynamics treatment in any
sense. They do not know anything about the dynamics of the full
theory. The key point now is that though being in some sense
``kinematical'', the coherent states for the Maxwell field are
nevertheless a very good starting point to compute approximate quantum
corrections as testified by the computations in the framework of
quantum optics \cite{QuantumOptics}. Certainly this analogy is not 
complete in that QED is equipped with a true Hamiltonian (rather than just 
a Hamiltonian constraint) but it shows nevertheless that sometimes 
kinematical states lead to rather good approximations.

In the present work we will proceed in the same spirit: We will not
seek states which are solutions to the constraint \textit{and}
approximately correspond to some classical geometry, but rather start by 
considering \textit{kinematical} semiclassical states.  

Consequently, we treat the Hamiltonians of the matter fields 
\textit{not} as pieces of the Hamiltonian constraint
but rather as observables. In particular, there will be no lapse function,
these Hamiltonian operators are simply {\it different} operators (from the 
Hamiltonian constraint operator). Although we will follow essentially
the steps performed in \cite{Thiemann:1998rt}, the fact that we are 
dealing with different operators allows us to change the quantization 
procedure slightly, for instance, the Hamiltonian operators  leave the 
cylindrical subspaces of ${\cal H}_{kin}$ separately invariant.  

It is hard to judge the validity of this approach as compared to the 
desirable full fledged solution of the Hamilton constraint. To shed
some light on this issue, in an appendix we consider a simple quantum 
mechanical model system. For this system we can show that the
expectation values of Dirac observables in coherent states  
states on the kinematical level numerically differ from the results of 
a treatment using dynamical coherent states, the differences are
tiny, however, as long as the energy of the system is macroscopic.

Another issue raised by the treatment outlined above is that
much depends on the choice of the state that is employed to play the
role of the semiclassical state. We will defer a discussion of this
fascinating topic to the companion paper
\cite{ST02} and only make some brief remarks here:

All candidate semiclassical states proposed so far are graph based states,
i.e. cylindrical functions in $\hilb{H}_{\text{kin}}$. Consequently,
this is assumed to be the case in the present work.  
The picture might however change substantially if  
ideas such as the averaging over infinitely many graph based states 
advocated in \cite{Bombelli:2000ua} could be employed. For some more
discussion on this point we refer to 
\cite{ST02,Thiemann:2002vj}.\\ 

Having adopted the above viewpoint on the Hamilton constraint, 
we can construct {\it approximate} $n-$particle Fock states 
propagating on fluctuating quantum spacetimes as follows: 
Denote by ${\cal M}$ the gravitational phase space of initial data on the 
hypersurface $\Sigma$ of the differentiable manifold $M=\Rl\times \Sigma$.
Let $m\in {\cal M}$ be initial data for some background spacetime.
An ordinary $n-$ particle state is an excitation of a vacuum state 
$\Omega^{\text{Fock}}_{\text{matter}}(m)$ in a usual Fock space 
${\cal H}^{\text{Fock}}_{\text{matter}}(m)$ which is of a completely 
different type than the background independent Hilbert space ${\cal 
H}^{\text{kin}}_{\text{matter}}$. The construction of 
that vacuum state (and the entire Fock space) {\it makes heavy 
use of the background metric 
in question}, here indicated by the explicit dependence of the state on 
the point $m$ in the gravitational phase space. This dependence slips in
because the state $\Omega^{\text{Fock}}_{\text{matter}}(m)$ is usually 
chosen as the 
ground state of some Hamiltonian operator $\hat{H}_{\text{matter}}(m)$
on the background spacetime in question. We now see what will 
heuristically happen 
when we start switching on the gravitational fluctuations as well:
{\it The dependence of $\Omega^{\text{Fock}}_{\text{matter}}(m)$ on $m$ 
has to become 
operator valued}! In other words, the vacuum state function 
$m\mapsto \Omega_{\text{matter}}(m)$ becomes a {\it vacuum operator}
$\hat{\Omega}:=\Omega_{\text{matter}}(\hat{m})$,
that is, a function of the matter degrees of freedom with values in 
${\cal L}({\cal H}^{\text{kin}}_{\text{grav}})\otimes 
{\cal H}^{\text{kin}}_{\text{matter}}$
where ${\cal H}^{\text{kin}}_{\text{matter}}$ is now necessarily 
a background independent matter Hilbert space (which we have chosen to 
be the one currently in use in QGR) and 
${\cal L}({\cal H}^{\text{kin}}_{\text{grav}})$ denotes the space of 
linear operators on a background independent geometry Hilbert space.
We see that the whole concept of an $n-$particle state becomes a very 
different one in the background independent context! Of course, we do not 
want a vacuum operator but a vacuum state on the full Hilbert space
${\cal H}^{\text{kin}}$ so that one will apply the vacuum operator 
to a state $\psi_{\text{grav}}(m)\in {\cal H}^{\text{kin}}_{\text{grav}}$, 
that is, $\Omega(m)=\hat{\Omega}\psi_m$. 

We conclude that a {\it 
fundamental} $n-$particle state of some matter type corresponding to
an ordinary $n-$particle state of the same matter type propagating on some
background spacetime described by the point $m$ in the gravitational 
phase space will be a complicated linear combination of states of the form
$\psi_{\text{grav}}(m)\otimes\psi_{\text{matter}} 
\in {\cal H}^{\text{kin}}_{\text{grav}}
\otimes {\cal H}^{\text{kin}}_{\text{matter}}\;.
$
How should this state be obtained from first principles? We propose the 
following strategy: Consider the full gravity coupled Hamiltonian 
operator $\hat{H}$ and construct a {\it annihilation operator} from it,
which is now an operator on the full Hilbert space ${\cal H}^{\text{kin}}$
and whose partial classical limit at the point $m$ of the 
gravitational phase space with respect to the gravitational 
degrees of freedom mirrors the usual Fock space annihilation operator
on the background spacetime described by $m$. 

This is what one should do. Now recall that the 
construction of Fock space annihilation operators on a given 
background involves, for instance, the construction of fractional powers 
of the Laplacian
operator on that background metric which is an operator in the 
one-particle (or first quantized matter) Hilbert space. Thus, our 
fundamental annihilation operator 
will involve a quantization of these Laplacian operators which therefore
become an operator on the tensor product of the one particle 
Hilbert space and the gravitational Hilbert space ${\cal 
H}^{\text{kin}}_{\text{grav}}$. While we are able to actually construct  
these operators in the present paper, as one can imagine, the formulas 
that we obtain are too complicated in order to do practical computations 
with present mathematical technology because fractional powers of the 
Laplacian are defined via its spectral resolution which is difficult 
to find. 

As an approximation to this exact computation we therefore 
propose to first compute the expectation value of the Laplacian 
operator in a gravitational coherent state and then to take its fractional 
powers. 
Now, {\it precisely} because we are using coherent states, this 
approximation will coincide with the exact calculation to zeroth order in 
$\hbar$ while for higher orders we presently do not know how significantly
results are changed. The details of these statements will be presented in 
section \ref{se4}.\\
 
The last step in the program is then to obtain, in principle testable, 
predictions from the theory obtained so far. For instance, we are 
interested in states of the form 
$\psi_{\text{grav}}(m)\otimes\psi_{\text{matter}} 
\in {\cal H}^{\text{kin}}_{\text{grav}}
\otimes {\cal H}^{\text{kin}}_{\text{matter}}
$
and wish then to construct an effective matter Hamiltonian operator as a 
quadratic form through the formula
$$
<\psi_{\text{matter}},\hat{H}^{\text{eff}}_{\text{matter}}(m)
\psi'_{\text{matter}}>_{{\cal H}^{\text{kin}}_{\text{matter}}}
:=
<\psi_{\text{grav}}(m)\otimes \psi_{\text{matter}},
\hat{H} \psi_{\text{grav}}(m)\otimes \psi'_{\text{matter}}
>_{{\cal H}^{\text{kin}}_{\text{grav}} \otimes 
{\cal H}^{\text{kin}}_{\text{matter}}}
$$
The operator $\hat{H}^{\text{eff}}_{\text{matter}}(m)$ already contains 
information about 
the quantum fluctuations of geometry. The quantum fluctuations of matter
are certainly much larger than those of geometry in the energy range of 
interest to us, however, as we are interested in Poincar\'e invariance 
violating effects, which are excluded by definition in ordinary QFT on 
Minkowski space, in order to study those we can neglect the quantum 
effects of matter as a first approximation (that is, we are dealing with
free field theories except for the coupling to the gravitational field). 
Therefore, we take the 
classical limit of $\hat{H}^{\text{eff}}_{\text{matter}}(m)$ 
and study the wave like solutions of the matter dynamics it generates.
One can also take the point of view that this procedure corresponds 
to the {\it first quantization of matter on a fluctuating spacetime}.
Second quantization will then be studied later on when we discuss 
$n-$particle states.

Adopting this viewpoint, as soon as a semiclassical state for the gravity 
sector
is chosen, translation and rotation symmetry is heavily broken on 
short
scales due to the discreteness of the underlying graph. The theory
describes fields propagating on random lattices, bearing a remarkable
similarity to models considered in lattice gauge theory
\cite{Christ:1982zq,Christ:1982ck,Christ:1982ci}. Due to the lack of
symmetry on short scales, notions such as plane waves and hence dispersion
relations can at best be defined in some large scale or low energy limit.
We will show that the problem of treating these limits is by no means
trivial and requires careful physical considerations. It is closely
related to the condensed matter physics problem of computing macroscopic
parameters of an amorphous (i.e. locally anisotropic and inhomogeneous)
solid from the parameters of its microscopic structure. 

To get a feeling for the
problem, we will sketch a one dimensional model system for
which we are able to find exact solutions. 
We will then turn to general
fields on random lattices and describe a procedure to obtain dispersion
relations valid in the long wavelength regime.\\

This concludes our explanatory exposition. We will now proceed to the 
details.
\section{Review of Quantum Kinematics of QGR}
\label{se3}
In QGR the manifold underlying spacetime is taken to be diffeomorphic with
$M=\R\times\Sigma$ where $\Sigma$ represents a 3d manifold of arbitrary 
topology. We will now summarize the essential aspects of the kinematical
framework of \cite{Thiemann:1998rq} for matter fields coupled to 
quantum gravity. We also introduce the Hamiltonians that we will be 
deriving dispersion relations for.

\teil{Gravity and Gauge Theory Sector}

The canonical pair consists of a $G$ connection $A^i_a(x)$ for 
a compact gauge group $G$ and a Lie$(G)$ valued  
densitized vector field $E_i^a(x)$ on $\Sigma$. Here we can treat 
all four interactions on equal footing. For the gravitational sector 
we have $G=SU(2)$ and   
the relation of the canonical pair to the classical ADM variables 
$q_{ab}$, $K_{ab}$ is 
\begin{equation*}
\det(q) q^{ab}=\iota E^a_iE^{bi},\qquad
A^i_a=\Gamma^i_a -\frac{\iota}{\sqrt{\det(q)}}K_{ab}E^{bi}, 
\end{equation*}
where $\Gamma$ is the spin connection corresponding to the triad $E$,
and $\iota$ is the \textit{Barbero-Immirzi} parameter which can in
principle take  any nonzero value in $\C$
\cite{Immirzi:1997dr,Barbero:1995an}. We will choose $\iota=1$  
in what follows. 
As for units, we choose $[A]=\text{meter}^{-1}$. As a consequence, 
$E$ will be dimensionless for gravity and has dimension cm$^{-2}$ for 
Yang-Mills theories.  

In the following we will have frequent opportunity to use the notion
of graphs embedded in $\Sigma$:
\begin{Definition}
By an \textit{edge} $e$ in $\Sigma$ we shall mean an equivalence class
of analytic maps $[0,1]\longrightarrow \Sigma$, where two such maps
are equivalent if they differ by an orientation preserving
reparametrization.\newline 
A \textit{graph} in $\Sigma$ is defined to be a set of edges such that 
two distinct ones intersect at most in their endpoints. 
\end{Definition}
There is some notation in connection to graphs that we will use
frequently:\newline 
The endpoints of an edge $e$ will be called \textit{vertices} and denoted by
$b(e)$ (the beginning point of $e$), $f(e)$ (the final point of $e$).
\newline
The set of edges of a graph $\gamma$ will be denoted by $E(\gamma)$,
the set of vertices of its edges 
by $V(\gamma)$.\newline
Given a graph $\gamma$, we will denote the edges of $\gamma$ having
$v$ as vertex by $E(\gamma,v)$ or $E(v)$ if it is clear which graph we 
are referring to.\newline
Given a graph $\gamma$, a vertex $v\in V(\gamma)$ 
and an edge $e\in E(v)$ we define
\begin{equation*}
  \sigma(v,e)=\begin{cases} 
+1& \text{ if } b(e)=v\\ 
-1&\text{ if }f(e)=v\\ 
0&\text{ if $e$ is not adjacent to $v$}\\
\end{cases}. 
\end{equation*}
Thus $e^{\sigma(v,e)}$ is always outgoing with respect to $v$. 

Being a one-form, $A$ can be integrated
naturally (that is, without recurse to background structure) along
piecewise analytic curves $e$ in $\Sigma$, to form \textit{holonomies}
\begin{equation*}
h_e[A]=\porder\exp\left[i\int_e A \right] \qquad \in G. 
\end{equation*}
It is convenient to consider a class of functionals of the connection
$A$ a bit more general:  
\begin{Definition}
A functional $f[A]$ of the connection is called \textit{cylindrical
  with respect to a piecewise analytical graph} $\gamma$ if there is a 
function
\begin{equation*}
f:G^{\betr{E(\gamma)}}\longrightarrow \C,
\end{equation*} 
such that 
\begin{equation}
\label{eq0.1}
  f[A]=f(h_{e_1}[A],h_{e_2}[A],\ldots),\qqquad e_1,e_2,\ldots\in
  E(\gamma). 
\end{equation}
\end{Definition} 
The density weight of $E$ on the other hand is such that, using an
additional real internal vector field $f^i$ it can be naturally integrated
over surfaces $S$ to form a quantity
\begin{equation*}
E_{S,f}= \int_S f^i\;(*E)_i
\end{equation*}
analogous to the electric flux through $S$.

In the connection representation of diffeomorphism invariant gauge field 
theory, quantization of the 
Poisson algebra generated by the classical 
functions $\cyl$ and the vector space space of electric fluxes $\E$ is 
achieved on the Ashtekar-Lewandowski Hilbert space
\begin{equation*} \label{3.0}
  \hilb{H}_0=\lzwo(\gcon,d\mu_0).
\end{equation*}
It is based on the compact Hausdorff space $\gcon$ of generalized
connections which is a suitable enlargement of the space of smooth 
connections $\con$ and the uniform measure $\mu_0$.

The classical Yang-Mills Hamiltonian (coupled to gravity) reads
\be \label{3.1}
H_{YM}=\frac{1}{2Q_{YM}}\int_\Sigma d^3x 
\frac{q_{ab}}{\sqrt{\det(q)}}[E^a_I E^b_J+B^a_I B^b_J] \delta^{IJ}.
\ee
Here $Q_{YM}$ is the Yang-Mills coupling constant, $E^a_I$ is the 
Yang-Mills electric field, $B^a_I=\epsilon^{abc} F_{bc}^I$ the magnetic
field associated with the Yang-Mills curvature $F_{ab}^I$. 

We now quantize this operator along the lines of 
\cite{Thiemann:1998rt}, 
actually only for Maxwell Theory since in this 
paper we are interested only in free theories when taking the metric as a 
background field.
Here we take advantage of the fact that $\hat{H}_{YM}$ is an operator of its 
own although, of course, the integrand of (\ref{3.1}) is a piece of the 
classical Hamiltonian constraint of geometry and matter. Accordingly we 
may exploit the quantization ambiguity concerning the loop attachment 
in \cite{Thiemann:1998rt} as follows: We define the operator 
$\hat{H}_{YM}$ consistently on the combined spin-network basis of matter 
and geometry introduced in \cite{Thiemann:1998rq} and use the following 
notion. 
\begin{Definition}
Let a graph $\gamma$, a vertex $v\in V(\gamma)$ and two different edges 
$e,e'\in E(\gamma)$ incident at $v$ be given. By a {\it minimal loop} 
based at $v$ we mean a loop $\beta(\gamma,v,e,e')$ in $\gamma$ which
\begin{itemize}
\item starts at $v$ along $e$ and ends at $v$ along $e'$,
\item does not self-overlap,
\item the number of edges used by $\beta$ except $e,e'$ cannot be reduced 
without breaking the loop into pieces.
\end{itemize}
\end{Definition}
Notice that given $\gamma,v,e,e'$ a minimal loop does not need to be unique!
Denote by $S(\gamma,v,e,e')$ the set of minimal loops corresponding to 
the data indicated and by $L(\gamma,v,e,e')$ their number.
Notice also that the notion of a minimal
loop does not make any reference to a background metric, it is an object 
that belongs to the field of {\it algebraic graph theory} \cite{AGT1,AGT2,AGT3}. 

Here we see the first difference as compared to the loop choice in 
\cite{Thiemann:1998rt}: A minimal loop is always contained in the 
graph that we are dealing with. 
The second difference that we will introduce in contrast to
\cite{Thiemann:1998rt} is that there we used functions of holonomies of the 
type 
$H_{\beta}-H_{\beta}^{-1}$ in order to express the Yang-Mills magnetic
field in terms of holonomies. However, as correctly pointed out in \cite{GR},
the regularization ambiguity allows more general functions, the only 
criterion is that the final operator is gauge invariant and that the 
function should vanish at trivial holonomy. 
Our preliminary proposal for  
the Maxwell Hamiltonian operator, projected to spin-network states over 
graphs $\gamma$ is then 
\ba \label{3.2}
\hat{H}_{M,\gamma} &=& -\frac{\alpha m_P}{2\ell_P^3} \sum_{v\in V(\gamma)}
\sum_{v\in e\cap e'} 
[\frac{3}{N(\gamma,v)}]^2
\hat{Q}^j_e(v,\frac{1}{2})\hat{Q}^j_{e'}(v,\frac{1}{2})
\times \nonumber\\
&\times & \{ -Y_e Y_{e'}+\frac{1}{P(\gamma,v,e)P(\gamma,v,e')\alpha^2}
[\sum_{v\in e_1\cap e_2;e_1,e_2\perp e} \frac{1}{L(\gamma,v,e_1,e_2)}
\sum_{\beta\in S(v,\gamma,e_1,e_2)} \frac{\ln(H_\beta)}{i}]
\times\nonumber\\
&& \times
[\sum_{v\in e'_1\cap e'_2;e'_1,e'_2\perp e'} \frac{1}{L(\gamma,v,e'_1,e'_2)}
\sum_{\beta\in S(v,\gamma,e'_1,e'_2)} \frac{\ln(H_\beta)}{i}]
\},
\ea
where for a vertex $v$, a real positive number $r$ and an edge $e$ 
starting at it we have defined the basic operator
\be \label{3.3}
\hat{Q}^j_e(v,r):=\frac{1}{4r}\mbox{tr}(\tau_j h_e[h_e^{-1},(\hat{V}_v)^r]),
\ee
with $\hat{V}_v$ the volume operator
\cite{Rovelli:1995ge,Ashtekar:1997fb} 
over an arbitrarily small open region 
containing $v$. We are using a basis of $su(2)$ with 
$\mbox{tr}(\tau_j\tau_k)=-2\delta_{jk}$.
Here $\alpha=\hbar Q_M$ is the Feinstruktur constant and $m_P,\ell_P$ are 
the Planck mass and length respectively. We have adapted the 
coefficients of \cite{Thiemann:1998rt} to 
the case of $G=U(1)$ and we have distinguished the Maxwell holonomy $H$
from the gravitational holonomy $h$. $Y_e$ is the right invariant vector 
field on $U(1)$ with respect to the degree of freedom $H_e$.
The notation $e\perp e'$ means that
$e\not=e',e\cap e'=v\not=\emptyset$ and that the tangents of $e,e'$ at 
$v$ are linearly independent. The number $P(\gamma,v,e)$ is the number
of pairs of edges $e_1,e_2$ with $v\in e_1\cap e_2;e_1,e_2\perp e$
and $N(\gamma,v)$ denotes the valence of the vertex $v$.
For each vertex $v$ the edges incident at $v$ are supposed to be outgoing
from $v$, otherwise replace $e$ by $e^{\sigma(v,e)}$ everywhere in 
(\ref{3.2}). The branch of the logarithm involved in 
(\ref{3.2}) is defined by $\ln(1)=0$. The logarithm is convenient in order
to define photon states later on but any other choice will do as well,
just giving rise to more quantum corrections.

The manifestly gauge invariant and spatially diffeomorphism invariant 
Hamiltonian (\ref{3.2}) is 
preliminary because we may want to order it differently later on.
Notice that it does not have the correct classical limit on an arbitrary
graph, the graph has to be sufficiently fine in order to reach it!
We will show that it defines a positive definite, essentially
self-adjoint operator on ${\cal H}_{\text{kin}}$.
\begin{Lemma}
For any positive real $r$ the operator
$i \widehat{Q}^j_e(v,r)$ on 
$\hilb{H}_0$ defined by (\ref{3.3})
is essentially self-adjoint with core given by the core of 
$\widehat{V}_v$.
\end{Lemma}
\begin{proof}
Since $(h_e)_{AB}$ is a bounded operator it suffices to show that
$i\widehat{Q}^j_e(v,r)$ is symmetric with dense domain the core 
of $\widehat{V}_v$.\newline
Using that $[(h_e)_{AB}]^\dagger=(h_e^{-1})_{BA}$ 
and $\cc{(\tau_J)_{AB}}=-(\tau_J)_{BA}$
we find
\begin{align*}
4r [i\;\widehat{Q}^j_e(v,r)]^\dagger 
&= i\cc{(\tau_j)_{AB}} 
\comm{\left((h_{e}^{-1})_{CA}\right)^\dagger}{V_v^r}
\left((h_{e})_{BC}\right)^\dagger
= -i(\tau_j)_{BA} 
\comm{\left((h_{e})_{AC}\right)}{V_v^r}
\left((h_{e}^{-1})_{CB}\right)\\
&= -i\tr\left(\tau_j 
\comm{\left((h_{e})\right)}{\widehat{V}_v^r}h_{e}\right)
= -i\tr\left(\tau_J 
h_{e}\widehat{V}_v^r
h_e^{-1}\right)\\
&= i\tr\left(\tau_j h_{e}
\comm{h_{e}^{-1}}{V_v^r}\right)
=4r [i\widehat{Q}^j_e(v,r)]
\end{align*}
because $\tr(\tau_j)\widehat{V}_v=0$.
\end{proof}

\teil{Scalar and Higgs Fields}

We will consider only Lie$(G)$ valued Higgs fields $\phi^i$ with 
canonically conjugate momentum $\pi_i$. In particular, a neutral
scalar field $\phi$ is Lie$(U(1))$ valued and transforms in the trivial 
adjoint representation. We will take $\phi_i$ to be dimensionless,
then $\pi_i\propto \dot{\pi}^i$ has dimension cm$^{-1}$.

The background independent Hilbert space of \cite{Thiemann:1998rq}
is based on the quantities
\be \label{3.3a}
U(x):=e^{\phi^i(x)\tau_i} \mbox{ and } \pi_{R,f}=\int_R d^3x f^i \pi_i,
\ee
where $U(x)$ is referred to as point holonomy, $\tau_j$ is a basis 
of Lie$(G)$ and  the second quantity
is diffeomorphism covariant since $\pi_i$ is a scalar density. 
One can then quantize the Poisson algebra generated by these objects 
on a Hilbert space $L_2(\overline{{\cal U}},d\mu_U)$ where 
$\overline{{\cal U}}$ is a distributional extension of the space 
$\cal U$ of smooth point holonomies and $d\mu_U$ is an associated 
uniform measure. This Hilbert space is very similar in spirit to the 
one for gauge theories displayed in (\ref{3.0}). A dense subspace
of functions in this Hilbert space consists of the cylindrical functions.
Here a function is cylindrical over a graph $\gamma$, if it depends only
on the point holonomies $U(v),\;v\in V(\gamma)$. That the point 
holonomies are restricted to the vertices of a graph is dictated by gauge 
invariance (for neutral scalar fields there is clearly no such argument
but given a function depending on a finite number of point holonomies
we can always trivially extend it to depend trivially on gravitational
holonomies over a graph with the arguments of the point holonomies as 
vertices).

In this paper we are only interested in neutral Klein-Gordon fields without 
interaction potential. The unitary operator $\hat{U}(x)$ acts by 
multiplication while the momentum operator is densely defined by
$$
\hat{\pi}_R f_\gamma=i\hbar Q_{KG} \sum_{v\in V(\gamma)\cap R} Y_v f_\gamma,
$$ 
where $Y_v$ denotes the right invariant vector field on $U(1)$ with respect
to the degree of freedom $U(v)$ and $Q_{KG}$ is the Klein Gordon
coupling constant which is such that $\hbar Q_{KG}$ has dimension cm$^2$.

The classical Klein Gordon Hamiltonian coupled to geometry is given by
\be \label{3.4}
H_{KG}=\frac{1}{2 Q_{KG}} \int_\Sigma d^3x[\frac{\pi^2}{\sqrt{\det(q)}}
+\sqrt{\det(q)}[q^{ab} \phi_{,a} \phi_{,b}+K^2 \phi^2]],
\ee
where $K^{-1}$ is the Compton wave length of the Klein Gordon field.
To quantize (\ref{3.4}) we again copy the procedure of 
\cite{Thiemann:1998rt} and define it on combined matter -- geometry 
spin-network states with the 
following modifications: 1) no new Higgs vertices on edges of the graph 
are introduced and 2) we replace the function 
$[U(x)-1 ]/i$ that substitutes $\phi(x)$ by something else in accordance 
to what we have said for gauge fields already.
We then propose the preliminary version of the Klein Gordon Hamiltonian 
operator, projected to spin-network states over graphs $\gamma$ by 
\ba \label{3.5}
\hat{H}_{KG,\gamma}&=&
-\frac{\hbar Q_{KG}}{2\ell_P^{11}} m_p\sum_{v\in V(\gamma)} Y_v^2 
\times\nonumber\\
&\times &
[\frac{1}{T(\gamma,v)}\sum_{v\in e_1\cap e_2\cap e_3;e_1\perp 
e_2\perp e_3}\frac{1}{3!} \epsilon_{ijk}\epsilon^{IJK} 
\hat{Q}^i_{e_I}(v,\frac{1}{2})\hat{Q}^j_{e_J}(v,\frac{1}{2})
\hat{Q}^k_{e_K}(v,\frac{1}{2})]^\dagger 
\times \nonumber\\
&&\times 
[\frac{1}{T(\gamma,v)}\sum_{v\in e'_1\cap e'_2\cap e'_3;e'_1\perp 
e'_2\perp e'_3}\frac{1}{3!} \epsilon_{lmn}\epsilon^{KLM} 
\hat{Q}^l_{e_L}(v,\frac{1}{2})\hat{Q}^m_{e_M}(v,\frac{1}{2})
\hat{Q}^n_{e_N}(v,\frac{1}{2})]
\nonumber\\
&+&
\frac{1}{2\hbar Q_{KG} \ell_P^7} m_p\sum_{v\in V(\gamma)} 
\times \nonumber\\
&\times&
[\frac{1}{2 T(\gamma,v)}\sum_{v\in e_1\cap e_2\cap e_3;e_1\perp e_2\perp e_3}
\epsilon^{IJK} \epsilon_{jkl}
\frac{[\ln(U(f(e_I))-\ln(U(b(e_I))]}{i}
\hat{Q}^k_{e_J}(v,\frac{3}{4})
\hat{Q}^l_{e_K}(v,\frac{3}{4})]^\dagger 
\times \nonumber\\
&&\times 
[\frac{1}{2 T(\gamma,v)}\sum_{v\in e'_1\cap e'_2\cap e'_3;e'_1\perp 
e'_2\perp e'_3} \epsilon^{LMN} \epsilon_{jmn}
\frac{[\ln(U(f(e_L))-\ln(U(b(e_L))]}{i}
\hat{Q}^m_{e_M}(v,\frac{3}{4})
\hat{Q}^n_{e_N}(v,\frac{3}{4})]
\nonumber\\
&+&
\frac{(K\ell_P)^2}{2\ell_P \hbar Q_{KG}} m_p\sum_{v\in V(\gamma)} 
[\frac{\ln(U(v)}{i}]^\dagger[\frac{\ln(U(v)}{i}]\hat{V}_v,
\ea
where $T(\gamma,v)$ is the number of triples of edges incident at $v$
with linearly independent tangents there and $b(e)$ and $f(e)$ respectively
denote starting point and end point of an edge. The operator (\ref{3.5})
is again manifestly gauge and diffeomorphism invariant. Notice that 
$\hbar Q_{KG}/\ell_P^2$ is dimensionless while $\hat{V}_v$ has dimension
cm$^3$ so that all terms have mass dimension. We already have ordered 
terms in (\ref{3.5}) in a manifestly positive way and the branch of the 
logarithm used corresponds to the fundamental domain of $\Cl$ again, 
i.e. $\ln(z/|z|)\in (-\pi,\pi]$.\\

\teil{Fermion Fields}

Since the 
canonical, non-perturbative quantization of the Einstein-Dirac 
theory in four spacetime dimensions using 
{\it real valued connections} is maybe less familiar to the reader
we review the essential aspects from \cite{Thiemann:1998rq} in 
slightly more detail. The classical Hamiltonian reads (we neglect
coupling to the Maxwell field in this paper since we want to isolate 
effects of quantum gravity on the propagation of free fields, see 
\cite{Thiemann:1998rq,Thiemann:1998rt} for the coupling to 
non-gravitational forces)
\be \label{3.10}
H=\hbar\int_\Sigma d^3x \{\frac{E^a_j}{2}[{\cal D}_a J_j 
+i(\bar{\psi}^T\sigma_j{\cal D}_a\psi-\bar{\eta}^T\sigma_j{\cal D}_a\eta
-c.c.)-K_a^j(\bar{\psi}^T\psi-\bar{\eta}^T\eta)]
+i K_0\sqrt{\det(q)}(\bar{\psi}^T\eta-\bar{\eta}^T\psi)\}
\ee
where $K_0$ is the rest frame wave number, $\sigma_j$ are the  
Pauli matrices, $J_j:=\bar{\psi}^T\sigma_j\psi+\bar{\eta}^T\sigma_j\eta$
is the fermion current, $\psi=(\psi^A)$ and $\eta=(\eta_{A'})$ respectively 
denote the 
left -- and right handed components of the Dirac spinor $\Psi=(\psi,\eta)^T$,
$\bar{\psi}$ denotes the involution on Gra{\ss}mann variables and the 
complex
conjugation $c.c.$ is meant in this sense.
The spinors $\psi,\eta$ transform as scalars under diffeomorphisms and as
left and right handed spinors under $SL(2,\Cl)$. In particular,
${\cal D}_a\psi=\partial_a\psi+\frac{1}{2}A_a^j\tau_j\psi,
{\cal D}_a\eta=\partial_a\eta+\frac{1}{2}A_a^j\tau_j\eta$ where 
$\tau_j=-i\sigma_j$. Our convention for the Minkowski space Dirac matrices 
is $\gamma^0=-i\sigma_2\otimes 1_2,\;\gamma^j=
\sigma_1\otimes\sigma_j$ appropriate for signature $(-,+,+,+)$.
The dimension of our spinor fields is cm$^{-3/2}$ so that (\ref{3.10})
has indeed dimension of energy. Notice the explicit appearance of the 
field $K_a^j=A_a^j-\Gamma_a^j$.

A peculiarity of spinor fields is that they are their own canonical 
conjugates. Consider the {\it half-densities} 
\be \label{3.11}
\xi:=\sqrt[4]{\det(q)}\psi,\;\;
\rho:=\sqrt[4]{\det(q)}\eta,
\ee
then the canonical anti-brackets are given by
\be \label{3.12}
\{\xi_A(x),\bar{\rho}_B(y)\}_+=\frac{\delta_{AB}\delta(x,y)}{i\hbar},\;\;
\{\rho_A(x),\bar{\rho}_B(y)\}_+=\frac{\delta_{AB}\delta(x,y)}{i\hbar},
\ee
while all other anti-brackets vanish. That (\ref{3.11}) mixes gravitational
and spinor degrees of freedom is absolutely crucial: without this peculiar
mixture it would not be $A_a^j$ that is canonically conjugate to 
$E^a_j/\kappa$ but rather $A_a^j+i\ell_p^2
e_a^j(\bar{\psi}^T\psi+\bar{\eta}^T\eta)$ 
which is now {\it complex valued} and this would destroy the 
Ashtekar Lewandowski Hilbert
space ${\cal H}_0$ since connections would become complex valued.

Clearly, we obtain the Einstein-Weyl Hamiltonian by setting either 
$\xi$ or $\rho$ to zero. Likewise we can treat the case of several
fermion species by adding appropriate similar terms to (\ref{3.10}).
In what follows we just stick with (\ref{3.10}), the reader may introduce
the appropriate changes for the case by hand himself.

In order to quantize (\ref{3.10}) we want to write 
(\ref{3.10}) into a more suggestive form. To that end,
notice that the Gauss constraint in the presence of fermions reads 
\be \label{3.13}
\frac{1}{\kappa} {\cal 
D}_a\frac{E^a_j}{\sqrt{\det(q)}}+\frac{\hbar}{2}J_j=0,
\ee
where $\kappa$ is the gravitational constant and 
where ${\cal D}_a$ acts on tensorial indices by the Christoffel connection
associated with $q_{ab}$ and on $SU(2)$ indices by the connection $A_a^j$. 
Thus we can solve the Gauss constraint for the 
fermion current so that after an integration by parts we have the 
identity
\be \label{3.14}
\hbar\int_\Sigma d^3x \frac{E^a_j}{2}[{\cal D}_a J_j] 
=\frac{1}{\kappa}\int_\Sigma d^3x \frac{({\cal D}_a E^a_j)^2}{\sqrt{\det(q)}}
\ee
modulo the Gauss constraint which now depends only on the gravitational
degrees of freedom. Clearly, in flat space (\ref{3.14}) vanishes.

Next it is easy to see that
\be \label{3.15}
+iE^a_j(\bar{\psi}^T\sigma_j{\cal D}_a\psi
-\bar{\eta}^T\sigma_j{\cal D}_a\eta-c.c.)
=i\frac{E ^a_j}{\sqrt{\det(q)}}(\bar{\xi}^T\sigma_j{\cal D}_a\xi 
-\bar{\rho}^T\sigma_j{\cal D}_a\rho-c.c.),
\ee
where ${\cal D}_a\xi:=\partial_a\xi+A_a^j\tau_j\xi/2$ ignores the  
density weight of $\xi$ (and similar for $\rho$) since the appropriate 
correction term is cancelled through a similar term in the $c.c.$ piece.

Formulas (\ref{3.14}) and (\ref{3.15}) imply that (\ref{3.10}) can be 
rewritten in terms of $\xi,\rho$ as 
\ba \label{3.15a}
H&=&\frac{1}{\kappa}\int_\Sigma d^3x 
\frac{({\cal D}_a E^a_j)^2}{\sqrt{\det(q)}}
\\
&+&
\hbar\int_\Sigma d^3x \{\frac{E^a_j}{2\sqrt{\det(q)}}[
+i(\bar{\xi}^T\sigma_j{\cal D}_a\xi-\bar{\rho}^T\sigma_j{\cal D}_a\rho
-c.c.)-K_a^j(\bar{\xi}^T\xi-\bar{\rho}^T\rho)]
+i K_0(\bar{\xi}^T\rho-\bar{\rho}^T\xi)\}.
\nonumber\\
\ea

Now recall from \cite{Thiemann:1998rq} in that for reasons of diffeomorphism
covariance it turned out to be crucial to work instead of with 
the half densities $\xi,\rho$ with the scalars
\be \label{3.16}
\theta_A(x):=\int_\Sigma d^3y \sqrt{\delta(x,y)} \xi_A(y)\;\;,\;\;
\theta'_A(x):=\int_\Sigma d^3y \sqrt{\delta(x,y)} \rho_A(y),
\ee
which still transforms covariantly under gauge transformations
$\theta(x)\to g(x)\theta(x)$ since the distribution 
$\sqrt{\delta(x,y)}$ has support at $x=y$. If we require the fields
$\theta$ to be ordinary Gra{\ss}mann fields, then formula 
(\ref{3.16}) implies that the spinor half-densities $\xi,\rho$ 
are distributional Gra{\ss}mann fields. This distributional character
is due to the factor $\root[4]\of{\det(q)}$ which in quantum theory
becomes an operator valued distribution proportional to
$\sqrt{\delta(x,y)}$ (recall that there is no such thing as classical
fermion fields). The inversion of (\ref{3.16}) is given
by
\be \label{3.17}
\xi_A(x):=\sum_{y\in\Sigma} \sqrt{\delta(x,y)} \theta_A(y)\;\;,\;\;
\rho_A(x):=\sum_{y\in\Sigma} \sqrt{\delta(x,y)} \theta'_A(y),
\ee
due to the identity 
$\sqrt{\delta(x,y)\delta(x,z)}=\delta(x,y)\delta_{y,z}$ where 
$\delta_{x,y}$ denotes the Kronecker symbol (equal to one when 
$x=y$ and zero otherwise).

Let $f_\epsilon(x,y)=f_\epsilon(y,x)=f_\epsilon(x-y)$ be a one parameter 
family of smooth, nowhere negative 
functions of rapid decrease such that also $\sqrt{f_\epsilon(x,y)}$
is smooth and such that $\lim_{\epsilon\to 0} f_\epsilon(x,y)=
\delta(x,y)$. An example would be 
$f_\epsilon(x,y)=\prod_a\;[e^{-(x^a-y^a)^2/(2\epsilon)}/\sqrt{2\pi\epsilon}]$.
Then 
\ba \label{3.18}
(\partial_a\theta)(x)&:=&
\lim_{\epsilon\to 0}\int_\Sigma d^3y (\partial_{x^a}\sqrt{f_\epsilon(x,y)}) 
\xi(y)
\nonumber\\
&=& -
\lim_{\epsilon\to 0}\int_\Sigma d^3y (\partial_{y^a}\sqrt{f_\epsilon(x,y)}) 
\xi(y)
\nonumber\\
&=& 
\lim_{\epsilon\to 0}\int_\Sigma d^3y \sqrt{f_\epsilon(x,y)} 
(\partial_a\xi)(y)
\nonumber\\
&=& 
\int_\Sigma d^3y \sqrt{\delta(x,y)} (\partial_a\xi)(y)
\ea
where in the integration by parts no boundary term was picked up since 
$f_\epsilon$ is of rapid decrease. It follows that
\be \label{3.19}
({\cal D}_a\theta)(x)=
\int_\Sigma d^3y \sqrt{\delta(x,y)} ({\cal D}_a\xi)(y)\;
\Rightarrow
({\cal D}_a\xi)(x)=
\sum_y \sqrt{\delta(x,y)} ({\cal D}_a\theta)(y)
\ee
for classical (smooth) $A_a^j$.

The Fermion Hilbert space now is constructed by means of Berezin integral
techniques where our basic degrees of freedom are the 
$\theta_A(x),\theta'_A(x)$ and their involutions. We start with only one 
Gra{\ss}mann degree of freedom and denote by
$\cal S$ superspace with anticommuting Gra{\ss}mann coordinates 
$\theta,\bar{\theta}$, that is, $\theta^2=\bar{\theta}^2=0,\; 
\theta\bar{\theta}=-\bar{\theta}\theta$.

A ``holomorphic'' function depends only on $\theta$ and not on 
$\bar{\theta}$ and is of the general form 
\be \label{3.20}
f(\theta)=a+b\theta
\ee
with arbitrary complex valued coefficients $a,b$ while a generic function
on $\cal S$ is of the general form
\be \label{3.21}
F(\bar{\theta},\theta)=a+b\theta+c\bar{\theta}+d\bar{\theta}\theta
\ee
with arbitrary complex valued coefficients $a,b,c,d$. The integral
of $F$ over $\cal S$ with respect to the ``measure'' 
$d\bar{\theta} d\theta$ is given by
\be \label{3.22}
\int_{\cal S} d\bar{\theta} d\theta F(\bar{\theta},\theta)=d.
\ee
A quantization of the canonical anti-brackets 
\be \label{3.23}
\{\theta,\theta\}_+=\{\bar{\theta},\bar{\theta}\}_+=0,\;\;
\{\bar{\theta},\theta\}_+=\{\theta,\bar{\theta}\}_+=\frac{1}{i\hbar}
\ee
and of the reality conditions
\be \label{3.24}
\overline{\theta}=\bar{\theta},\;\;\overline{\bar{\theta}}=\theta
\ee
can be given on the space $L_2({\cal S},d\mu_F)$
of ``square-integrable'' holomorphic functions with respect to the 
``probability measure'' 
\be \label{3.25}
d\mu_F:=e^{\bar{\theta}\theta}\;d\bar{\theta}\; d\theta= 
[1+\bar{\theta}\theta]\; d\bar{\theta}\; d\theta 
\ee
which is positive definite:
\be \label{3.26}
<f,f'>:=\int_{\cal S} d\mu_F(\bar{\theta},\theta)
\overline{f(\theta)}f'(\theta)=\bar{a}a'+\bar{b}b' \mbox{ for }
f(\theta)=a+b\theta,\;f'(\theta)=a'+b'\theta.
\ee
We just need to define the operators $\hat{\theta},\hat{\bar{\theta}}$ 
by
\be \label{3.27}
(\hat{\theta}f)(\theta):=\theta f(\theta)=a\theta,\;
(\hat{\bar{\theta}}f)(\theta):=\frac{d}{d\theta} 
f(\theta)=b
\ee
(derivative from left)
and verify immediately that the canonical anticommutation relations 
\be \label{3.28}
[\hat{\theta},\hat{\theta}]_+=2\hat{\theta}^2
=[\hat{\bar{\theta}},\hat{\bar{\theta}}]_+=2\hat{\bar{\theta}}^2=0,\;\;
[\hat{\bar{\theta}},\hat{\theta}]_+=[\hat{\theta},\hat{\bar{\theta}}]_+=
\hat{\bar{\theta}}\hat{\theta}+\hat{\theta}\hat{\bar{\theta}}=1
\ee
as well as the adjointness relations 
\be \label{3.29}
\hat{\theta}^\dagger=\hat{\bar{\theta}},\;\;
\widehat{\bar{\theta}}^\dagger
=\hat{\theta}
\ee
hold with respect to the measure $d\mu_F$.

This covers the quantum mechanical case. Let us now come to the 
case at hand. Recall that we had the following anti-brackets for 
our spinor degrees of freedom
\be \label{3.30}
\{\xi_A(x),\xi_B(y)\}_+=\{\bar{\xi}_A(x),\bar{\xi}_B(y)\}_+=0,\;\;
\{\xi_A(x),\bar{\xi}_B(y)\}_+=\{\bar{\xi}_B(y),\xi_A(x)\}_+=
\frac{\delta_{AB}\delta(x,y)}{i\hbar},
\ee
and similar for $\rho$. Inserting the transformation (\ref{3.16})
we see that (\ref{3.30}) is equivalent with
\be \label{3.31}
\{\theta_A(x),\theta_B(y)\}_+=\{\bar{\theta}_A(x),\bar{\theta}_B(y)\}_+=0,\;\;
\{\theta_A(x),\bar{\theta}_B(y)\}_+=\{\bar{\theta}_B(y),\theta_A(x)\}_+=
\frac{\delta_{AB}\delta_{x,y}}{i\hbar},
\ee
so the $\delta$ distribution is simply replaced by the Kronecker symbol.
This suggests to define the Fermion Hilbert space as the {\it continuous}
infinite tensor product \cite{Thiemann:2000by}
\be \label{3.32}
{\cal H}^\otimes_D:=\otimes_{x\in\Sigma,A=\pm 1/2} L_2({\cal S},d\mu_F)
\ee
where $\hat{\theta}_A(x),\hat{\bar{\theta}}_A(x)\equiv
\hat{\theta}_A(x)^\dagger$ are densely defined on $C_0$ vectors by
\ba \label{3.33}
\hat{\theta}_A(x)\otimes_f&:=&
[\otimes_{x\not=y,B} f_{y,B}]\otimes[f_{x,-A}\otimes(\hat{\theta} f_{x,A})],
\nonumber\\
\hat{\theta}_A(x)^\dagger\otimes_f&:=&
[\otimes_{x\not=y,B} f_{y,B}]\otimes[f_{x,-A}\otimes(\hat{\theta}^\dagger 
f_{x,A})].
\ea
This Hilbert space is unnecessarily large for the following reason: 
Due to gauge invariance the spinor fields are confined to the vertices of 
an at most countably infinite graph. In particular, if we are dealing 
with finite graphs only, then the subspace ${\cal H}_D$ of 
the Hilbert space (\ref{3.32}), defined as the inductive limit of the 
cylindrical spaces
\be \label{3.34}
{\cal H}_{\gamma,D}
:=\otimes_{v\in V(\gamma),A=\pm 1/2} L_2({\cal S},d\mu_F)
\ee
via the isometric monomorphisms $\hat{U}_{\gamma\gamma'}$ for 
$\gamma\subset\gamma'$, densely by
\be \label{3.35}
\hat{U}_{\gamma\gamma'}:\;{\cal H}_{\gamma,D}\mapsto {\cal H}_{\gamma',D}
;\;\otimes_f^\gamma:=\otimes_{v\in V(\gamma),A} f_{v,A}
\mapsto [\otimes_{v\in V(\gamma),A} f_{v,A}]\;
[\otimes_{x\in V(\gamma')-V(\gamma),B} 1]
\ee
is completely sufficient for our purposes in this paper (as long as 
$\sigma$ is 
compact, otherwise we can use the techniques from \cite{Thiemann:2000by}). Equation 
(\ref{3.35})
displays ${\cal H}_D$ as the strong equivalence class Hilbert subspace
of ${\cal H}^\otimes_D$ formed by the $C_0$ vector $1:=\otimes_{x,A} 1$. 

We now turn to the quantization of (\ref{3.15a}). Actually we will not 
consider the terms which are proportional to $K_a^j,({\cal D}_a E^a_j)^2$
because they vanish in flat space (with which we are mainly concerned
in this paper). Of course, quantum corrections will give non-vanishing
corrections but since we are doing only exploratory calculations in this 
paper, let us just not discuss those terms. Then the methods of 
\cite{Thiemann:1998rq} lead to the the following quantum operator
restricted to matter -- geometry spin network functions over a graph 
$\gamma$
\ba \label{3.36}
&&\hat{H}_{D,\gamma}=-\frac{m_P}{2\ell_P^3}\sum_{v,v'\in V(\gamma)}
[\hat{\theta}_B(v')\hat{\theta}^\dagger_A(v)
-\hat{\theta}'_B(v')\hat{\theta}^{\prime\dagger}_A(v)]
\times \nonumber\\
&\times&
\{\{ \frac{1}{T(\gamma,v)} \epsilon_{ijk} \epsilon^{IJK}
\sum_{v\in e_1\cap e_2\cap e_3;e_1\perp e_2\perp e_3}
\hat{Q}^i_{e_I}(v,\frac{1}{2})\hat{Q}^j_{e_J}(v,\frac{1}{2})
[\tau^k (h_{e_K}\delta_{v',f(e_K)}-\delta_{v',b(e_K)})]_{AB}\}
\nonumber\\
&&-
\{ \frac{1}{T(\gamma,v')} \epsilon_{ijk} \epsilon^{IJK}
\sum_{v'\in e_1\cap e_2\cap e_3;e_1\perp e_2\perp e_3}
[(h_{e_K}^{-1}\delta_{v,f(e_K)}-\delta_{v,b(e_K)})\tau^k]_{AB}
\hat{Q}^i_{e_I}(v',\frac{1}{2})\hat{Q}^j_{e_J}(v',\frac{1}{2})\}
\}
\nonumber\\
&&
-i\hbar K_0\sum_{v,v'\in V(\gamma)}
\delta_{AB}\delta_{v,v'} [\hat{\theta}'_B(v')\hat{\theta}^\dagger_A(v)
-\hat{\theta}_B(v')\hat{\theta}^{\prime\dagger}_A(v)].
\ea
It is not difficult to see that this operator is self-adjoint. Again,
as compared to \cite{Thiemann:1998rt} we have chosen a different ordering 
and there are no new fermion vertices created on cylindrical functions
over $\gamma$.\\

\teil{Remark}

We have defined the operators $\hat{H}_M,\hat{H}_{KG},\hat{H}_D$
in the combined spin-network basis of matter and 
geometry defined in \cite{Thiemann:1998rq}. Such a spin network
function $T_s(A,A_M,\phi,\theta,\theta')$ carries a label
$s=(\gamma,\vec{j},\vec{n},\vec{m},\vec{B},\vec{B}',\vec{I})$ consisting 
of a graph $\gamma$, a coloring of its edges $e$ with Einstein 
non-zero spins 
$j_e\in\vec{j}$ and non-zero Maxwell charges $n_e\in \vec{n}$ as well as a 
coloring of its vertices $v$ by non-zero scalar charges $m_v\in \vec{m}$, 
non-zero left-handed fermion helicities $B_v\in \vec{B}$,  
non-zero right-handed fermion helicities $B'_v\in \vec{B}'$ and 
intertwiners
$I_v\in \vec{I}$ which make the state gauge invariant under the action of 
the gauge group $SU(2)\times U(1)$. This defines densely a continuum 
operator by $\hat{H}T_s:=\hat{H}_{\gamma(s)} T_s$ on the 
Hilbert space ${\cal H}={\cal H}_E\otimes
{\cal H}_M\otimes{\cal H}_{KG}\otimes{\cal H}_D$ ($E$ stands for the 
Einstein sector) where 
$\hat{H}=\hat{H}_M+\hat{H}_{KG}+\hat{H}_D$. However, the operators
$\hat{H}_\gamma$ are not the cylindrical projections of $\hat{H}$
since they are not cylindrically consistent, i.e.
$\hat{H}_{\gamma}\mbox{Cyl}_{\gamma'}\not=     
\hat{H}_{\gamma'}\mbox{Cyl}_{\gamma'}$ for $\gamma'\subset\gamma$.
Rather, in order to evaluate the operator on cylindrical functions, one 
has to decompose them in terms of spin-network functions. Nevertheless
one can construct from the family $(\hat{H}_\gamma)$ a cylindrically
consistent family $(\hat{H}^\gamma)$ as follows:\\

Notice that each ${\hat H}_\gamma$ has the following structure
\be \label{3.37}
\hat{H}_\gamma=\sum_{v\in V(\gamma)} \hat{H}_{\gamma,v},
\ee
where $\hat{H}_v$ is a local operator, that is, it depends only 
on the finite subset $E_v(\gamma)\subset E(\gamma)$ of edges of $\gamma$ 
incident at $v$. For $e\in E(\gamma)$ denote by $\hat{P}_e$
the projection operator on the closed linear span of cylindrical
functions in Cyl$_\gamma$ which depend through non-zero spin on the 
edge $e$. For subsets $E\subset E'\subset E(\gamma)$ let 
\be \label{3.38}
\hat{P}_{\gamma,E,E'}:=
[\prod_{e\in E} \hat{P}_{\gamma,e}]\;[\prod_{e'\in E'-E} 
(1-\hat{P}_{\gamma,e'})].
\ee
Consider now the operator
\be \label{3.39}
\hat{H}^\gamma_v:=\sum_{E\subset E_v(\gamma)}
\hat{P}_{\gamma,E,E_v(\gamma)}\hat{H}_{\gamma-[E_v(\gamma)-E],v}
\hat{P}_{\gamma,E,E_v(\gamma)}
\ee
where the sum is over the power set of $E_v(\gamma)$ (set of all
subsets) and with it the, still self-adjoint, cylindrically 
consistent family of operators 
\be \label{3.40}
\hat{H}^\gamma=\sum_{v\in V(\gamma)} \hat{H}^\gamma_v.
\ee
Notice that
$\hat{H}_{\gamma-[E_v(\gamma)-E],v}\equiv 0$ whenever $|E|<3$.

It is important to notice that both families $(\hat{H}^\gamma)$ 
and $(\hat{H}_\gamma)$ give rise to the same continuum operator as long as 
there are no gravitational holonomy operators outside of 
commutators involved (as is the case for 
the bosonic pieces). It is just sometimes more convenient to have 
a consistent operator family if one does not want to decompose a 
cylindrical function into spin-network functions. In fact, our 
semiclassical states over $\gamma$ are 
not spin-network states over $\gamma$ but they are cylindrical 
functions, i.e. linear combinations of spin-network states 
where also all smaller graphs $\gamma'\subset\gamma$ appear. 

The careful reader will rightfully ask whether the expectation values of
the operators defined in terms of (\ref{3.37}) and (\ref{3.40}) 
respectively agree on semiclassical states 
over $\gamma$ (they do exactly if no gravitational holonomy operator is 
involved). This is important since the operator 
$\hat{P}_{\gamma,E,E_v(\gamma)}$ did not come out of the derivation in 
\cite{Thiemann:1998rt} (for operators involving the gravitational 
holonomy) and thus could spoil the classical limit. Fortunately,
the answer to the question is affirmative due to two reasons:\\
1) The expectation value
of $\hat{P}_{\gamma,e}$ turns out to be of the form $1-e^{-c/t^\beta}$ where 
$c,\beta$ are positive numbers of order unity {\it for non-degenerate 
metrics} and $t$ is a tiny number related to $\hbar$. Then the expectation 
value of $1-\hat{P}_{\gamma,e}$ is of order $O(t^\infty)$.\\
2) We will use only graphs for semiclassical calculations such 
that the valence of the vertices is bounded from above. Thus the sum
(\ref{3.39}) involves only a small number of terms and 
$(1-e^{-c/t^\beta})^{|E_v(\gamma)|}=1+O(t^\infty)$.\\
Thus the expectation
value of $\hat{H}_\gamma$ with respect to cylindrical functions over
$\gamma$ agrees with that
of $\hat{H}^\gamma$ to {\it any finite order in $t$} and for 
semiclassical calculations we can practically treat the family 
$(\hat{H}_\gamma)$ as if it was cylindrically consistently defined.  
We will assume that to be the case in what follows.
\section{Matter $n-$Particle States on Graphs}
\label{se4}
The aim of this section is to sketch how one would in principle construct
exact $n-$particle states propagating on fluctuating quantum geometries as 
well approximations of those  by using quantum geometry expectation values 
in gravitational coherent states. The computation of those expectation 
values is sketched in the next section, more details 
can be found in our companion paper \cite{ST02}. Since in these two 
papers we are only 
interested in qualitative features we do not want to spend too much 
technical effort and therefore make our life simple by replacing $SU(2)$
by $U(1)^3$, see \cite{Thiemann:2000ca,Thiemann:2000bx} for how 
non-Abelian gauge groups blow up the computational effort by an order of 
magnitude. We leave the exact computation for future investigations
after the conceptual issues discussed in this work have been 
settled.\footnote{Actually,  
$SU(2)$ is replaced by $U(1)^3$ in the $G_{Newton}\to 0$ limit if one 
rescales the gravitational connection $A$ by $A/G_{Newton}$ 
(I\"on\"u-Wigner contraction), but $G_{Newton}\to 0$ also implies
$\ell_P\to 0$ and this is precisely the regime we are interested in. 
However, ultimately we must do the $SU(2)$ computation.}

\subsection{Specialization to Cubic Random Graphs}
\label{s4.1}

A second simplification that we will make is to consider for the 
remainder of this and the companion paper only graphs of cubic topology.
Graphs of different topology can be treated in principle by the same 
methods that we develop below but for analytical computations graphs 
of different topology present an extremely hard book keeping problem
which is presumably only controllable on a computer. But apart from 
these more practical considerations we can also give some physical 
motivation:\\
A)\\
As is well-known, there exist an infinite number of discretizations
of a classical continuum action or Hamiltonian with the correct continuum
limit and some of them reflect the continuum properties of the action or
Hamiltonian better than others. In that respect it is relevant to mention
the existence of so-called {\it perfect actions} \cite{perfect1} 
which arise as fixed points of the renormalization group flow for Euclidean 
field theories. These are 
perfect in the sense that although one works at the discretized level,
the expectation values are, for instance, {\it Euclidean invariant}!
These techniques have been applied also to differential operators
and there exist, for instance, Euclidean invariant Laplace operators 
on arbitrarily coarse cubic lattices \cite{perfect2,perfect3} despite the fact that
a cubic lattice seems to introduce an unwanted direction dependence!
The quantization of our Hamiltonian operators could exploit that freedom
in order to improve semiclassical properties,
the choice that we have made is merely a first natural guess.\\
B)\\
Secondly, the graph does not need to 
be ``regular" but rather could be an oriented {\it random} cubic graph
adapted to the three metric $q_{ab}$ to be approximated. In fact, we will
discuss this possibility in detail in our companion paper when we discuss
(light) propagation on random cubic graphs.
Such a graph could be obtained by a suitable random process. 
Let us sketch a procedure for two dimensional Euclidean space:  
We start with a sequence of randomly chosen
vectors $\evec{v}_1,\evec{v}_2,\ldots$ 
subject to the condition that the angle between vectors adjacent in
the sequence lies in the range $[-\pi/2,\pi/2]$.   
Then a first sequence of vertices of the graph could be obtained as 
$0,\evec{v}_1, \evec{v}_1+\evec{v}_2,\ldots$ and a first sequence of
edges by connecting the vertices by straight lines. 
Now choose another sequence of vectors subject to the same condition
on the angles, as well as an additional vector $\evec{p}$. 
Again we obtain a set of vertices
$\evec{p},\evec{p}+\evec{v}_1,\evec{p}+\evec{v}_1+\evec{v}_2,\ldots$ 
and corresponding edges. These are to be discarded if an edge
intersects one of the edges obtained before. Otherwise the vertices and 
edges are added to the graph. This can be iterated until a convenient
number of vertices has been obtained. Then the rest of the edges
necessary to turn the graph into a cubic one can be obtained by
connecting the vertices ``vertically'' by straight lines. 
This description is certainly sketchy, but it could easily be made
precise by supplying the details of the random distributions for the
choice of the points, directions etc. Similarly, it can be
generalized to three dimensions and curvilinear edges. 

\begin{figure}
\centerline{\epsfig{file=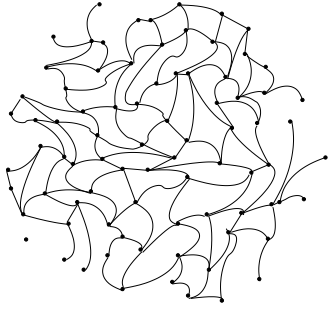, height=4cm}}
\caption{\label{fi9} Part of a (two dimensional) random graph of cubic 
  topology.}
\end{figure}

Although it would be extremely cumbersome to obtain analytical results 
about the resulting random graphs, procedures like the one sketched above 
can be implemented on a computer in a straightforward way and any 
desired information can then be obtained numerically.  
An artistic impression of a small part of a random graph of cubic
topology is given in figure \ref{fi9}. Note that it does not favor a 
direction on a large scale, though it is certainly not rotationally 
symmetric in a strict sense.  

A random cubic graph is certainly still diffeomorphic to a regular
cubic graph in $\Rl^3$ with its natural Cartesian orientation. Its vertices 
$v$ can be thought of as points in $\Zl^3$, its edges can be labelled
as $e_I(v),\;I=1,2,3$ with $b(e_I(v))=v,\;f(e_I(v))=v+I$ where 
$v+I$ denotes the next neighbor vertex of $v$ along the $I-$direction.
Given a 3-metric $q_{ab}$ to be approximated, let $\epsilon$ be the 
average length of the $e_I(v)$ with respect to it. Notice that the angles
and shapes of the edges are completely random! Depending on the random
process that has generated the cubic graph, within each close to flat
coordinate patch there will also be an isotropy scale $\delta=N\epsilon$ at 
which the the graph looks homogeneous and isotropic. The meaning of the 
scales $\epsilon,\delta$ with respect to physical processes such as light 
propagation which introduces a length scale of its own, namely the 
wave length $\lambda$, is subject to a discussion in our companion paper
where we will see how these scales fit together with the quantum gravity 
scale $\ell_P$ and what their relative sizes should be. For the purposes 
of this section we just need that the graph $\gamma$ in question has 
cubic topology.\\
\\
We can now specialize our formulas for the family of matter Hamiltonians 
$(\hat{H}_\gamma)$ to cubic graphs $\gamma$. Given $v\in V(\gamma)$
let $e^+_I(v):=e_I(v),\; e^-_I(v):=(e_I(v-I))^{-1}$ so that 
$b(e^\pm_I(v))=v$. Then it is easy to check that the local volume operator
$\hat{V}_v$ becomes
\be \label{4.1}
\hat{V}_v=\ell_P^3 \sqrt{|\epsilon^{jkl}
[\frac{Y^{e^+_1(v)}_j-Y^{e^-_1(v)}_j}{2}]\;
[\frac{Y^{e^+_2(v)}_k-Y^{e^-_2(v)}_k}{2}]\;
[\frac{Y^{e^+_3(v)}_l-Y^{e^-_3(v)}_l}{2}]|}
\ee
where $Y^e_j$ denotes the right invariant vector field on $SU(2)$ with 
respect to the degree of freedom $h_e$. Notice that we used the coefficient 
$1/(8\cdot 3!)$ that was derived in
\cite{Rovelli:1995ge,Ashtekar:1997fb}. As we will see, with 
respect to coherent states of the type constructed in 
\cite{Thiemann:2000ca,Thiemann:2000bx}, {\it only cubic graphs will assign
correct expectation values to the volume operator if the coefficient 
$1/(8\cdot 3!)$ is used}!

For a graph of cubic topology we have 
$P(\gamma,v,e)=4,\;T(\gamma,v)=8$ since each vertex is 
6-valent. Also, there are 12 minimal loops based at $v$ along the 
edges $e^{\sigma_1}_I(v),e^{\sigma_2}_J(v),\;I\not=J$ each of which is 
unique, that is $L(\gamma,v,e,e')=1$, 
due to our simple lattice topology. Given $v,I,\sigma$ there are 4 minimal
loops $\beta^I_{\sigma;\sigma_1,\sigma_2}(v)$ along the
edges $e^{\sigma_1}_J(v),e^{\sigma_2}_K(v)$ with $\epsilon_{I,J,K}=1$
whose orientation we choose to be such that the tangents of
$e^\sigma_I(v),e^{\sigma_1}_J(v),e^{\sigma_2}_K(v)$ at $v$ in this order 
form a $3\times 3$ matrix of positive determinant.
With the notation $h_I^\sigma(v):=h_{e_I^\sigma(v)},\;\sigma=\pm 1$
and $\hat{Q}^j_{I\sigma}(v,r):=\hat{Q}^j_{e^\sigma_I(v)}(v,r)$
our matter Hamiltonian constraint operators (\ref{3.2}), (\ref{3.5})
and (\ref{3.36}) become, respectively:\\
\\
{\bf Maxwell Field:}
\ba \label{4.2}
\hat{H}_{M,\gamma} &=& -\frac{\alpha m_P}{2\ell_P^3} \sum_{v\in V(\gamma)}
\sum_{I,J,\sigma,\sigma'} 
\hat{Q}^j_{I\sigma}(v,\frac{1}{2})\hat{Q}^j_{J\sigma'}(v,\frac{1}{2}) \times 
\nonumber\\
&\times & \{- \frac{1}{4} Y_{I\sigma}(v) 
Y_{J\sigma'}(v)+\frac{1}{64\alpha^2}
[\sum_{\sigma_1,\sigma_2} 
\frac{\ln(H_{\beta^I_{\sigma;\sigma_1,\sigma_2}(v)})}{i}]\;
[\sum_{\sigma'_1,\sigma'_2} 
\frac{\ln(H_{\beta^J_{\sigma';\sigma'_1\sigma'_2}(v)})}{i}]
\}
\ea
where $Y_{I\sigma}(v)=Y_{e^\sigma_I(v)}$.\\
Remark:\\
A remark is in order concerning the logarithms that appear in (\ref{4.2}).
Recall that in the previous section we defined for a loop $\beta$ the 
number $\ln(H_\beta)$ by the main branch of the logarithm, specifically
$\ln(H_\beta)/i\in[-\pi,\pi)$. Now for a minimal loop $\beta$ and sufficiently
fine $\gamma$ it is indeed true, for a classical connection, that
$\ln(H_\beta)=\ln(H_{e_1})+..+\ln(H_{e_N})$ if $\beta=e_1\circ..\circ e_N$
where all appearing logarithms are with respect to the same branch.
Notice that the right hand side is gauge invariant only for small
gauge transformations.
In the case of a cubic graph we will show below that this relation can be
written, e.g. $\ln(H_{\beta^I_{1;1,1}(v)})=
\epsilon^{IJK} \partial^+_J\ln(H_{e_K(v)}$ where $(\partial^+_I f)(v)=
f(v+I)-f(v)$. Using the transversal projector $P^\perp$ 
of (\ref{4.16}) we can further
write this as 
$\ln(H_{\beta^I_{1;1,1}(v)})=
\epsilon^{IJK} P^\perp\cdot\partial^+_J
P^\perp\cdot \ln(H_{e_K(v)}$. Finally, again
for sufficiently fine $\gamma$ we may write this as 
$\ln(H_{\beta^I_{1;1,1}(v)})=
\epsilon^{IJK} \partial^+_J
[(P^\perp\cdot \ln(H_{e_K(v)})(\mbox{mod} 2\pi)]$. The quantity in the 
square bracket is now manifestly gauge invariant even after extending to
distributional connections. It is this definition that we will be 
using for $\ln(H_{\beta^I_{1;1,1}(v)})$ in what follows, it simply 
corresponds to a different choice for the quantization ambiguity
(for cubic graphs).

When we compute commutators of such logarithms with electric flux
operators on sufficiently fine graphs 
we must regularize the commutator by {\it first} restricting to
classical connections (so that the value of the logarithm lies in 
$(-\pi,\pi)$ and {\it then} extending the result to 
$\ab$. This being understood we get $[Y_{I\sigma}(v),\ln(H_\beta)]=
H_\beta^{-1}[Y_{I\sigma}(v),H_\beta]$ without picking up discontinuities.
Similar remarks hold for the scalar field.\\
\\
{\bf Klein-Gordon Field}:
\ba \label{4.3}
\hat{H}_{KG,\gamma}&=&
-\frac{\hbar Q_{KG}}{2\ell_P^{11}} m_p\sum_{v\in V(\gamma)} Y_v^2 
\times\nonumber\\
&\times &
[\frac{1}{8}\sum_{\sigma_1,\sigma_2,\sigma_3} 
\frac{\sigma_1\sigma_2\sigma_3}{3!} \epsilon_{ijk}\epsilon^{IJK}
\hat{Q}^i_{I,\sigma_1}(v,\frac{1}{2})\hat{Q}^j_{J,\sigma_2}(v,\frac{1}{2})
\hat{Q}^k_{K,\sigma_3}(v,\frac{1}{2})]^\dagger
\times\nonumber\\ &&\times
[\frac{1}{8}\sum_{\sigma'_1,\sigma'_2,\sigma'_3} 
\frac{\sigma'_1\sigma'_2\sigma'_3}{3!} \epsilon_{lmn} \epsilon^{LMN}
\hat{Q}^l_{L,\sigma'_1}(v,\frac{1}{2})\hat{Q}^m_{M,\sigma'_2}(v,\frac{1}{2})
\hat{Q}^n_{N,\sigma'_3}(v,\frac{1}{2})]
\nonumber\\
&+&
\frac{1}{2\hbar Q_{KG} \ell_P^7} m_p\sum_{v\in V(\gamma)} 
\times \nonumber\\
&\times&
[\frac{\epsilon^{IJK}}{16} \epsilon_{jkl}
\sum_{\sigma_1,\sigma_2,\sigma_3}
\frac{[\sigma_1\partial_{\sigma_1,I}\ln(U)](v)}{i}
\hat{Q}^k_{J\sigma_2}(v,\frac{3}{4})
\hat{Q}^l_{K\sigma_3}(v,\frac{3}{4})]
\times\nonumber\\ &&\times
[\frac{\epsilon^{LMN}}{16} \epsilon_{jmn}
\sum_{\sigma'_1,\sigma'_2,\sigma'_3}
\frac{[\sigma'_1\partial_{\sigma'_1,I}\ln(U)](v)}{i}
\hat{Q}^m_{M\sigma'_2}(v,\frac{3}{4})
\hat{Q}^n_{N\sigma'_3}(v,\frac{3}{4})]
\nonumber\\
&+&
\frac{(K\ell_P)^2}{2\ell_P \hbar Q_{KG}} m_p\sum_{v\in V(\gamma)} 
[\frac{\ln(U(v)}{i}]^\dagger[\frac{\ln(U(v)}{i}]\hat{V}_v
\ea
where for a function $F:\;V(\gamma)\mapsto \Cl$ we have defined the 
{\it edge derivative} $[\partial_e F](v):=F(f(e))-F(v)$ if  
$v=b(e)$. Specialized to the cubic graph we write 
$(\partial_{\sigma,I} F)(v):=(\partial_{e^\sigma_I(v)} F)(v)$
and 
$(\partial^\sigma_I F)(v):=\sigma(\partial_{\sigma,I} F)(v)$
is the {\it forward (backward) edge derivative} at $v$ if $\sigma=1$
($\sigma=-1$).\\
{\bf Dirac Field}:
\ba \label{4.4}
&& \hat{H}_{D,\gamma}=-\frac{m_P}{2\ell_P^3}\sum_{v,v'\in V(\gamma)}
[\hat{\theta}_B(v')\hat{\theta}^\dagger_A(v)
-\hat{\theta}'_B(v')\hat{\theta}^{\prime\dagger}_A(v)]
\times \nonumber\\
&\times&
\{ \frac{1}{8} \epsilon_{ijk} \epsilon^{IJK}
\sum_{\sigma_1,\sigma_2,\sigma_3}
\hat{Q}^i_{I\sigma_1}(v,\frac{1}{2})\hat{Q}^j_{J\sigma_2}(v,\frac{1}{2})
[\tau^k 
(h^{\sigma_3}_K(v)\delta_{v',f(e^{\sigma_3}_K(v))}-\delta_{v',v})]_{AB}\}
\nonumber\\
&&-
\{ \frac{1}{8} \epsilon_{ijk} \epsilon^{IJK}
\sum_{\sigma_1',\sigma_2',\sigma_3'}
[([h^{\sigma_3'}_K(v')]^{-1}\delta_{v,f(e^{\sigma_3'}_K(v'))}-
\delta_{v,v'})\tau^k]_{AB}
\hat{Q}^i_{I\sigma_1'}(v',\frac{1}{2})\hat{Q}^j_{J\sigma_2'}(v',\frac{1}{2})\}
\}
\nonumber\\
&&
-i\hbar K_0\sum_{v,v'\in V(\gamma)}
\delta_{AB}\delta_{v,v'} [\hat{\theta}'_B(v')\hat{\theta}^\dagger_A(v)
-\hat{\theta}_B(v')\hat{\theta}^{\prime\dagger}_A(v)].
\ea
The operator (\ref{4.4}) is not manifestly positive definite on a 
non-flat background. Fortunately, in contrast to the bosonic 
Hamiltonians, positivity is not required in order to arrive at
suitable annihilation operators since it is already normally ordered
(subject to the usual particle -- antiparticle reinterpretation upon
passage to second quantization). 
In order to obtain positivity of this Hamiltonian we have to invoke a 
positive and negative energy decomposition of the 1-particle Hilbert
space as done in QFT on CST. 
This step will not be preformed here as it goes beyond the exploratory 
purposes of this paper.
\subsection{One-Particle Hilbert Spaces on a Graph}
\label{s4.2}
The operators in (\ref{4.2}), (\ref{4.3}) and (\ref{4.4}) define operators 
on the Hilbert space ${\cal H}_{kin}=
{\cal H}^E_{kin}\otimes{\cal H}^M_{kin}\otimes{\cal 
H}^{KG}_{kin}\otimes{\cal H}^D_{kin}$ with spin-network projections of
the form
\be \label{4.6}
\hat{H}_\gamma=\sum_{v,l;v',l'} 
\hat{M}_l(v)^\dagger_l(v) \hat{G}_{(v,l);(v',l')} \hat{M}_{l'}(v'),
\ee
where $v,v'\in V(\gamma)$ and $l,l'$ are elements of a discrete 
label set ${\cal L}$ of  
labels like the labels $j,I,\sigma,A,\mu$ where $\mu=1,2$ and 
$\theta^1=\theta,\theta^2=\theta'$. $\hat{M}_l(v)$ is a linear matter 
operator for each pair $(v,l)$ while $\hat{G}_{(v,l),(v',l')}$ is a 
geometry operator for each pair of pairs $(v,l),(v',l')$.
Our aim is to reorder $\hat{H}_\gamma$ in such a way that it acquires 
the usual form in terms of creation and annihilation operators. 
The Fermion Hamiltonian is already in this desired form, because 
$$
[\hat{\theta}^\alpha_A(v),\hat{\theta}^\beta_B(v')]_+=
\delta^{\alpha\beta}\delta_{AB}\delta_{vv'}
$$
already satisfies the canonical anticommutation relations but the bosonic 
terms do not. In order to do this we need to introduce the
{\it one particle Hilbert spaces} ${\cal H}^1_\gamma$ on the graphs $\gamma$.
These are defined as spaces of complex valued functions
$F:\;V(\gamma)\times {\cal L} \to \Cl;\;(v,l)\mapsto F_l(v)$ which are 
square summable, that is, their norm with respect to the inner product
\be \label{4.7}
<F,F'>_{{\cal H}^1_\gamma}:=\sum_{v,l} \overline{F_l(v)} F'_l(v)
\ee
converges. Thus, ${\cal H}^1_\gamma=\ell_2(V(\gamma)\otimes {\cal L})$
is equipped with a counting measure. Next to this we also introduce 
the Hilbert subspaces 
${\cal H}^E_\gamma,{\cal H}^M_\gamma, {\cal H}^{KG}_\gamma, 
{\cal H}^D_\gamma$ of square integrable cylindrical functions over $\gamma$
of the respective field type.
Having done this, we can consider 
the gravitational operator 
$\hat{G}_{(v,l),(v',l')}\in {\cal L}({\cal H}^E_\gamma)$ (${\cal L}(.)$ 
denotes the space of linear operators over $(.)$) as an operator 
$\hat{G}\in {\cal L}({\cal H}^E_\gamma\otimes{\cal H}^1_\gamma)$
densely defined by
\be \label{4.8}
[\hat{G} T^E_s\otimes F](A,(v,l)):=\sum_{(v',l')} 
[\hat{G}_{(v,l),(v',l')} T^E_s](A)\; F_{l'}(v'),
\ee
where $T^E_s$ is a gravitational spin network state over $\gamma$ and 
$A\in \ab_E$ is a gravitational generalized connection.
That the right hand side of (\ref{4.8}) is indeed again an $L_2$ function 
will be shown below. 

Likewise, we may consider the operators 
$\hat{M}_l(v)\in {\cal L}({\cal H}^{\text{matter}}_\gamma)$ as operators
$\hat{M}:\;{\cal H}^{\text{matter}}_\gamma\to
{\cal H}^{\text{matter}}_\gamma\times {\cal H}^1_\gamma$
densely defined by
\be 
[\hat{M} T^{\text{matter}}_s](A,(v,l)):=
[\hat{M}_l(v)T^E_s](A) 
\ee
where $T^{\text{matter}}_s$ is a matter spin network state.

But even better than that, for the bosonic pieces of 
(\ref{4.7}) we will be able to show that {\it the operator $\hat{G}$
is positive definite}! By inspection then, the whole (bosonic 
piece of the) operator $\hat{H}_\gamma$ is a positive operator.
Moreover, we will be able to take square roots of this operator, defined
in terms of its spectral resolution on the Hilbert space 
${\cal H}^E_\gamma\otimes{\cal H}^1_\gamma$. These square roots are 
precisely those that one would take on the Hilbert space 
${\cal H}^1_\gamma$ if the gravitational field was a background field
in order to arrive at the annihilation and creation operator 
decomposition.\\
\\
Let us now proceed to the details:\\
\\
{\bf Maxwell Hamiltonian}\\
\\
The electromagnetic Gau{\ss} constraint operator applied to cylindrical
functions over $\gamma$ reads \cite{Thiemann:2001yy}
\be \label{4.9}
\widehat{\mbox{Gau{\ss}}}(\Lambda)_\gamma=\sum_{v\in V(\gamma)}\Lambda(v)
[\sum_{e\in E(\gamma);b(e)=v} Y_e-\sum_{e\in E(\gamma);f(e)=v} Y_e].
\ee
Since we are working with gauge invariant functions, the Gauss constraint
is identically satisfied. Using the notation $Y^I(v):=Y_{e_I(v)}$ we 
obtain the operator identity
\be \label{4.10}
\sum_I [Y_{I,+}(v)+Y_{I,-}(v)]
=\sum_I [Y^I(v)-Y^I(v-I)]\equiv (\partial^-_I Y^I)(v)=0,
\ee
where naturally the backward edge derivative has popped out (we used 
$Y_{e^{-1}}=-Y_e$) and Einstein's summation convention is implicit. 
 
Next consider the loops $\beta^I_{\sigma_1;\sigma_2,\sigma_3}$.
It is easy to check that with $\epsilon_{IJK}=1$ for fixed $I$ we have
\be \label{4.11}
\ln(H_{\beta^I_{\sigma_1;\sigma_2,\sigma_3}}(v))=\sigma_1
[\ln(H_{e_J(v')})+\ln(H_{e_K(v'+J)})
-\ln(H_{e_J(v'+K)})-\ln(H_{e_K(v')}]_{v'=v+
\frac{\sigma_2-1}{2}J+\frac{\sigma_3-1}{2}K}.
\ee
Using the notation 
$\hat{A}_I(v):=[P^\perp\cdot\ln(H_{e_I(v)})/i]\mbox{mod}(2\pi)$ 
with $P^\perp$ defined in (\ref{4.16}) we 
can rewrite (\ref{4.11}) as (subject to the remark after (\ref{4.2})) 
\be \label{4.12}
\ln(H_{\beta^I_{\sigma_1;\sigma_2,\sigma_3}}(v))=i\sigma_1
\epsilon^{IMN}(\partial^+_M \hat{A}_N)_{v'=v+
\frac{\sigma_2-1}{2}J+\frac{\sigma_3-1}{2}K}
\ee
where naturally the forward edge derivative has appeared. Equation 
(\ref{4.12}) is obviously gauge invariant under 
$\hat{A}_I\mapsto \hat{A}_I+\partial^+_I F$. 

It is important to notice that forward and backward derivatives commute 
with each other, $[\partial^\sigma_I,\partial^{\sigma'}_J]=0$ for any
$I,J,\sigma,\sigma'$. We can now introduce the one-particle Hilbert space
${\cal H}^1_{M,\gamma}$ with inner product
\be \label{4.13}
<F,F'>_{{\cal H}^1_{M,\gamma}}=\sum_{v,I}\overline{F_I(v)} F'_I(v)
\ee
and one easily checks that 
$(\partial^\sigma_I)^\dagger=-\partial^{-\sigma}_I$ is the adjoint of the 
lattice derivative on ${\cal H}^1_{\gamma,M}$. 

It is convenient to introduce the lattice Laplacian
\be \label{4.14}
(\Delta f)(v):=\sum_I (\partial_I^-\partial_I^+ f)(v)
=\sum_I [f(v+b_I)+f(v-b_I)-2 f(v)]
\ee
which is easily seen to be negative definite on ${\cal H}_{\gamma,1}$
\be \label{4.15}
<F,\Delta F>_{{\cal H}_{\gamma,1}}=-\sum_{v\in V(\gamma)}
\sum_{I,J} |(\partial_J^- A_I)(v)|^2 
\ee
and invertible since ${\cal H}^1_{M,\gamma}$ does not contain zero 
modes by definition (they are not normalizable). With its help we may 
define the lattice transversal projector 
\be \label{4.16}
(P_\perp\cdot  F)_I(v)=F_I(v)-[\partial_I^+\frac{1}{\Delta}\partial_J^- 
F^J](v)=:\sum_{v',v} P^\perp_{(v,I),(v',J)} F^J(v').
\ee
We then obtain the operator identities
\be \label{4.17}
Y^I(v)=(P_\perp\cdot Y)^I(v) \mbox{ and }
\epsilon^{IJK}(\partial^+_J \hat{A}_K)(v)=
\epsilon^{IJK}(\partial^+_J [P_\perp \hat{A}]_K)(v).
\ee
It is important to realize that the lattice metric $\delta_{IJ}$ is not
a background structure, but is actually {\it diffeomorphism invariant},
it is the same for all cubic lattices and only depends on the topology
of the lattice (which in our case is cubic). Therefore the index position
of the index $I$ is actually irrelevant, in particular, $P_\perp=P^\perp$. 
Cubic graphs are distinguished by the fact the same projector $P_\perp$
in ${\cal H}^1_{M,\gamma}$ maps to the space of solutions to the Gauss
constraint $\partial_I^- F^I=0$ and to the gauge invariant piece of 
$F_I$ under $F_I\mapsto F_I+\partial^+_I f$.
Notice that indeed $P_\perp^2=P_\perp=P_\perp^\dagger,
\;\delta_{IJ}P_{IJ}^\perp(v,v')=2\delta_{v,v'}$
is a symmetric projector on ${\cal H}_{\gamma,1}$ 
on the two physical degrees of freedom per lattice point as 
desired. It is remarkable that all the structure that comes with 
$\partial_I^\pm$ {\it can be constructed without any reference to the 
gravitational degrees of freedom}! Of course, this is due to the fact that
the exterior derivative of a one form and the divergence of a vector 
density are metric independent.

Let us write the Hamiltonian $\hat{H}_{M,\gamma}$ in our new notation.
In order to simplify our life for the exploratory purposes of this paper 
we will replace the sum over the four loops corresponding to the 
choices $\sigma_1,\sigma_2$ divided by four in (\ref{4.2}) by one loop
corresponding to $\sigma_1=\sigma_2=1$. Likewise we replace the sum 
over the choices $\sigma$ divided by two by the term corresponding to 
$\sigma=1$. This just corresponds to the 
exploitation of the quantization ambiguity from which all the Hamiltonians
constructed so far suffer anyway. Then (\ref{4.2}) can be written in the 
compact form  
\ba \label{4.18}
\hat{H}_{M,\gamma} &=& -\frac{\alpha m_P}{\ell_P^3} \sum_{v\in V(\gamma)}
\hat{Q}^j_I(v,\frac{1}{2})\hat{Q}^j_I(v,\frac{1}{2}) \times 
\nonumber\\
&\times & \{ 
[-P_\perp\cdot Y]^I(v) [P_\perp\cdot Y]^I(v)+
[\epsilon^{IKL}\partial^+_K (P_\perp\cdot \hat{A})_L](v)
[\epsilon^{JMN}\partial^+_M (P_\perp\cdot \hat{A})_N](v)
\}
\ea
where $\hat{Q}^j_I(v,r)=\hat{Q}^j_{I,+}(v,r)$. Let 
\be \label{4.19}
\hat{Q}_{IJ}(v,r):=
[\hat{Q}^j_I(v,r)]^\dagger\hat{Q}^j_J(v,r)
=-\hat{Q}^j_I(v,r)\hat{Q}^j_J(v,r).
\ee
Notice that while the operators $\hat{Q}^j_I(v,r),\hat{Q}^j_J(v,r)$
do not commute, in (\ref{4.18}) only the the symmetric piece
of $\hat{Q}_{IJ}(v,\frac{1}{2})$ survives in (\ref{4.18}). We now define
$\hat{G}^{M;1}_{(v,I),(v',J)}, \hat{G}^{M;2}_{(v,I),(v',J)}$ as operators
on ${\cal H}^E_\gamma\otimes {\cal H}^1_{M,\gamma}$ by
\ba
&&[\hat{G}^{M,1}\psi\otimes F](A,(v,I)) := 
\sum_{v',K} P^\perp_{(v,I),(v',J)}
(\hat{Q}_{JK}(v',\frac{1}{2})\psi)(A) (P_\perp \cdot F)^K(v')
\nonumber\\
&& [\hat{G}^{M;2}\psi\otimes F](A,(v,I)) := 
-\sum_{v',K} P^\perp_{(v,I),(v',J)}\epsilon^{JKL} \partial^-_{v'K}
(\hat{Q}_{LM}(v',\frac{1}{2})\psi)(A) \epsilon^{MNP}
(\partial^+_N (P_\perp \cdot F)_P)(v').\nonumber
\ea
We can then write the Maxwell Hamiltonian in the even more compact form
\be \label{4.21}
\hat{H}^M_\gamma=\frac{\alpha m_P}{2\ell_P^3}
[<Y^\dagger,\hat{G}^{M;1}Y>_{{\cal H}^1_{M,\gamma}}
+<\hat{A}^\dagger,\hat{G}^{M;2}\hat{A}>_{{\cal H}^1_{M,\gamma}}]
\ee
where the adjoint in (\ref{4.21}) is with respect to ${\cal H}^M_\gamma$.
The following results are crucial.
\begin{Theorem} \label{th4.1} ~~~~\\
i)\\
The operators $\hat{G}^{M;1},\hat{G}^{M;2}$ are positive semidefinite
and definite on the subspace 
${\cal H}^E_\gamma\otimes {\cal H}^{1\perp}_{M,\gamma}$
where ${\cal H}^{1\perp}_{M,\gamma}=P_\perp\cdot {\cal 
H}^1_{M,\gamma}$.\\
ii)\\
The operator $\hat{H}^M_\gamma$ is positive definite on 
${\cal H}^E_\gamma\otimes {\cal H}^M_\gamma$ and thus 
$\hat{H}^M$ is positive definite on ${\cal H}$.
\end{Theorem}
Proof:\\
i)\\
Let $\Psi=\sum_\mu z_\mu \psi^E_\mu \otimes F_\mu \in 
{\cal H}^E_\gamma\otimes {\cal H}^1_{M,\gamma}$ be given. Then
\ba \label{4.22}
&& 
<\Psi,
\hat{G}^{M;1}\Psi>_{{\cal H}^E_\gamma \otimes {\cal H}^1_{M,\gamma}}
\nonumber\\
&=& \sum_{\mu,\nu} \bar{z}_\mu z_\nu
<\psi^E_\mu\otimes F_\mu,
\hat{G}^{M;1}\psi^E\otimes F_\nu>_{{\cal H}^E_\gamma \otimes {\cal 
H}^1_{M,\gamma}}
\nonumber\\
&=& \sum_{\mu,\nu} \bar{z}_\mu z_\nu \sum_{v,I,J}
\overline{(P\cdot F_\nu)^J(v)}
<\psi^E_\mu,\hat{Q}_{IJ}(v)\psi^E_\nu>_{{\cal H}^E_\gamma}
(P\cdot F_\nu)^J(v)
\nonumber\\
&=& \sum_v \sum_j 
||\sum_\mu z_\mu \sum_I (P\cdot F_\mu)^I(v) 
[\hat{Q}^j_I(v,\frac{1}{2})\psi^E_\mu]||^2_{{\cal H}^E_\gamma}
\ea
where we used that $P=P_\perp$ commutes with edge derivatives.
By the same manipulations we arrive at 
\ba \label{4.23}
&& 
<\Psi,
\hat{G}^{M;2}\Psi>_{{\cal H}^E_\gamma \otimes {\cal H}^1_{M,\gamma}}
\nonumber\\
&=& \sum_v \sum_j 
||\sum_\mu z_\mu \sum_I (P\cdot \partial^+\times F_\mu)^I(v) 
[\hat{Q}^j_I(v,\frac{1}{2})\psi^E_\mu]||^2_{{\cal H}^E_\gamma}
\ea
where $(\partial^\sigma\times F)^I=\epsilon^{IJK}\partial^\sigma_J F_K$.
The definiteness statement is clear by inspection.\\
ii)\\
Let now $\Psi^{EM}=\sum_\mu z_\mu \psi^E_\mu \otimes \psi^M_\mu \in 
{\cal H}^E_\gamma \otimes {\cal H}_{M,\gamma}$ be given. Then
\ba \label{4.24}
&& 
<\Psi^{EM},\hat{H}_{M,\gamma}\Psi^{EM}>_{{\cal H}^E_\gamma \otimes 
{\cal H}_{M,\gamma}}
\nonumber\\
&=& \sum_{\mu,\nu} \bar{z}_\mu z_\nu
<\psi^E_\mu\otimes \psi^M_\nu,
\hat{H}^M_\gamma \psi^E\otimes \psi^M_\nu>_{{\cal H}^E_\gamma 
\otimes {\cal H}_{M,\gamma}}
\nonumber\\
&=& \sum_{\mu,\nu} \bar{z}_\mu z_\nu \sum_{v,I,J}
<\psi^E_\mu,\hat{Q}_{IJ}(v)\psi^E_\nu>_{{\cal H}^E_\gamma}
[<(P\cdot Y)^I(v)\psi^M_\mu, 
(P\cdot Y)^I(v)\psi^M_\nu>_{{\cal H}_{M,\gamma}}
\nonumber\\
&& +<(P\cdot \partial^+\times \hat{A})^I(v)\psi^M_\mu, 
(P\cdot \partial^+\times Y)^J(v)\psi^M_\mu>_{{\cal H}_{M,\gamma}}]
\nonumber\\
&=& \sum_v \sum_j
||\sum_\mu z_\mu [\hat{Q}^j_I(v,\frac{1}{2})\psi^E_\mu]\otimes
[(P\cdot Y)^I(v)\psi^M_\mu]||^2_{{\cal H}^E_\gamma \otimes {\cal 
H}_{M,\gamma}}
\nonumber\\
&&+ \sum_v \sum_j
||\sum_\mu z_\mu [\hat{Q}^j_I(v,\frac{1}{2})\psi^E_\mu]\otimes
[(P\cdot \partial^+\times \hat{A})^I(v)\psi^M_\mu]||^2_{{\cal H}^E_\gamma 
\otimes {\cal H}_{M,\gamma}}.
\ea
The definiteness statement follows from the fact that we are working on 
the space of gauge invariant functions (solutions to the Gauss 
constraint).\\
$\Box$\\
It follows from this theorem that the operators 
$\hat{G}^{M;1},\hat{G}^{M;2}$ have an inverse on the subspace
${\cal H}^E_\gamma\otimes {\cal H}^{1\perp}_{M,\gamma}$. This fact 
is vital in order to arrive at the creation and annihilation operator 
decomposition of the Einstein-Maxwell Hamiltonian. We state here without
proof that the positivity property of $\hat{H}^{M}_\gamma$ holds on every 
graph so that it extends to the whole Hamiltonian $\hat{H}^M$. The 
latter statement follows from the fact that by construction the 
consistently defined Hamiltonian (\ref{3.40}) preserves the space of 
gravitational spin-network functions over a graph due to the projection 
operators on both sides in (\ref{3.39}). (Actually this is 
already true for the operator defined on spin-network functions 
because our operators do not change the graph on which a state 
depends). Therefore, if 
$\Psi=\psi^E_{\gamma_1}\otimes \psi^M_{\gamma_1}+\psi^E_{\gamma_1}\otimes 
\psi^M_{\gamma_1}$ we have 
\be \label{4.24a}
<\Psi,\hat{H}^M\Psi>_{{\cal H}_{kin}}=
<\psi^E_{\gamma_1}\otimes 
\psi^M_{\gamma_1},\hat{H}^M_{\gamma_1}
\psi^E_{\gamma_1}\otimes \psi^M_{\gamma_1}>
+
<\psi^E_{\gamma_2}\otimes 
\psi^M_{\gamma_2},\hat{H}^M_{\gamma_2}
\psi^E_{\gamma_2}\otimes \psi^M_{\gamma_2}>
\ee
due to orthonormality of the spin network functions and both terms are 
separately positive. \\
\\
{\bf Klein-Gordon Hamiltonian}\\
\\
Here the one-particle Hilbert space ${\cal H}^1_{KG,\gamma}$ is equipped 
with the inner product
\be \label{4.25}
\scpr{F}{F'}_{{\cal H}^1_{KG,\gamma}}=\sum_{v\in V(\gamma)} 
\overline{F(v)} F(v').
\ee
Let us also simplify the Klein-Gordon Hamiltonian (\ref{4.3}) by replacing 
the 
average over the eight right oriented triples of edges by a single one
corresponding to $\sigma_1=\sigma_2=\sigma_3=1$
\ba \label{4.26}
\hat{H}_{KG,\gamma}&=&
-\frac{\hbar Q_{KG}}{2\ell_P^{11}} m_p\sum_{v\in V(\gamma)} Y_v^2 
\times\nonumber\\
&\times &
[\frac{1}{3!} \epsilon_{ijk} \epsilon^{IJK}
\hat{Q}^i_I(v,\frac{1}{2})\hat{Q}^j_J(v,\frac{1}{2})
\hat{Q}^k_K(v,\frac{1}{2})]^\dagger \;
[\frac{1}{3!} \epsilon_{lmn}\epsilon^{LMN}
\hat{Q}^l_L(v,\frac{1}{2})\hat{Q}^m_M(v,\frac{1}{2})
\hat{Q}^n_N(v,\frac{1}{2})]
\nonumber\\
&+&
\frac{1}{2\hbar Q_{KG} \ell_P^7} m_p\sum_{v\in V(\gamma)} 
\times \nonumber\\
&\times&
[\frac{\epsilon^{IJK}}{2} \epsilon_{jkl}
\frac{[\partial^+_I\ln(U)](v)}{i}
\hat{Q}^k_J(v,\frac{3}{4})
\hat{Q}^l_K(v,\frac{3}{4})]^\dagger \;
[\frac{\epsilon^{LMN}}{2} \epsilon_{jmn}
\frac{[\partial^+_L \ln(U)](v)}{i}
\hat{Q}^m_M(v,\frac{3}{4})
\hat{Q}^n_N(v,\frac{3}{4})]
\nonumber\\
&+&
\frac{(K\ell_P)^2}{2\ell_P \hbar Q_{KG}} m_p\sum_{v\in V(\gamma)} 
[\frac{\ln(U(v)}{i}]^\dagger[\frac{\ln(U(v)}{i}]\hat{V}_v.
\ea
One can read off from (\ref{4.26}) the operators 
$\hat{G}^{KG;1}_{v,v'},\hat{G}^{KG;2}_{v,v'}$ on 
${\cal H}^E_\gamma$ giving rise to operators 
$\hat{G}^{KG;1},\hat{G}^{KG;2}$ on 
${\cal H}^E_\gamma\otimes {\cal H}^1_{KG,\gamma}$
which allows us to write (\ref{4.26}) in the compact form
\be \label{4.27}
\hat{H}^{KG}_\gamma=\frac{1}{2}
[<Y^\dagger,\hat{G}^{KG;1} Y>_{{\cal H}^1_{KG,\gamma}}
+<\hat{\phi}^\dagger,\hat{G}^{KG;1} \hat{\phi}>_{{\cal H}^1_{KG,\gamma}}]
\ee
where the the dagger is with respect to ${\cal H}^{KG}_\gamma$ and we used 
the notation $\hat{\phi}(x)=\ln(U(x))/i$. Of course there is no
projection operator involved in this case since in this paper we deal
with neutral scalar fields only.

A theorem analogous to theorem \ref{th4.1} can be proved in this case 
as well and will be left to the ambitious reader.\\
\\
{\bf Fermion Hamiltonian}\\
\\
Finally, let us also simplify (\ref{4.4}) by replacing the average over 
the eight triples of edges to the one corresponding to
$\sigma_1=\sigma_2=\sigma_3=1$, that is  
\ba \label{4.28}
&& \hat{H}_{D,\gamma}=-\frac{m_P}{2\ell_P^3}\sum_{v,v'\in V(\gamma)}
[\hat{\theta}_B(v')\hat{\theta}^\dagger_A(v)
-\hat{\theta}'_B(v')\hat{\theta}^{\prime\dagger}_A(v)]
\times \nonumber\\
&\times& \{
\{\epsilon_{ijk} \epsilon^{IJK}
\hat{Q}^i_I(v,\frac{1}{2})\hat{Q}^j_J(v,\frac{1}{2})
[\tau^k 
(h_K(v)\delta_{v',v+K}-\delta_{v',v})]_{AB}\}
\nonumber\\
&&-
\{\epsilon_{ijk} \epsilon^{IJK}
[([h_K(v')]^{-1}\delta_{v,v'+K}-
\delta_{v,v'})\tau^k]_{AB}
\hat{Q}^i_I(v',\frac{1}{2})\hat{Q}^j_J(v',\frac{1}{2})\}
\}
\nonumber\\
&&
-i\hbar K_0\sum_{v,v'\in V(\gamma)}
\delta_{AB}\delta_{v,v'} [\hat{\theta}'_B(v')\hat{\theta}^\dagger_A(v)
-\hat{\theta}_B(v')\hat{\theta}^{\prime\dagger}_A(v)]
\ea
where $h_I(v)=h_{e_I(v)}$.
One can read off from (\ref{4.28}) the operator
$\hat{G}^D_{(v,A,\mu),(v',B,\nu)}$
on ${\cal H}^E_\gamma$ giving rise to the operator 
$\hat{G}^D$ on 
${\cal H}^E_\gamma\otimes {\cal H}^1_{D,\gamma}$. Here $\mu=1,2$
where $\theta^1_A(v)=\theta_A(v)$ and $\theta^2_A(v)=\theta'_A(v)$
and ${\cal H}^1_{D;\gamma}$ is equipped with the inner product
\be \label{4.29}
<F,F'>_{{\cal H}^1_{D,\gamma}}=\sum_{v\in V(\gamma)} \sum_{A,\mu} 
\overline{F^\mu_A(v)} F^\mu_A(v')
\ee
This allows us to write (\ref{4.28}) in the compact form
\be \label{4.30}
\hat{H}^{D}_\gamma=
<\hat{\theta}^\dagger,\hat{G}^D \hat{\theta}>_{{\cal H}^1_{D,\gamma}}
\ee
where the the dagger is with respect to ${\cal H}^D_\gamma$.
Of course there is no
projection operator involved in this case since in this paper we deal
with neutral spinor fields only.

A theorem analogous to theorem \ref{th4.1} has not been proved in this 
case and is fortunately not necessary in order to arrive at creation 
and annihilation operators. See the remark at the end of subsection
\ref{4.2}.
\subsection{Fundamental Fock States and Normal Ordering}
\label{s4.3}
We notice that both bosonic Hamiltonians have the structure
\be \label{4.31}
\hat{H}_\gamma=\frac{1}{2}\sum_{v,v',l,l'} 
\hat{p}_l(v)\hat{P}((v,l),(v',l')) \hat{p}_{l'}(v') 
+\hat{q}_l(v)\hat{Q}((v,l),(v',l')) \hat{q}_{l'}(v'),
\ee
where $\hat{p}_l(v),\hat{q}_l(v)$ are operator valued distributions on 
${\cal H}_{matter,\gamma}$ with canonical commutation relations 
$[\hat{p}_l(v),\hat{q}_l(v')]=i\hbar \delta_{v,v'}\delta_{ll'}$ and
$\hat{P}((v,l),(v',l')),\hat{Q}((v,l),(v',l'))$ 
define positive definite operators on ${\cal H}_{geo,\gamma}\otimes 
{\cal H}^1_\gamma$. In particular, they are symmetric there, that is,
$\hat{P}((v,l),(v',l'))^\dagger=\hat{P}((v',l'),(v,l))$ where 
the dagger is with respect to ${\cal H}_{geo,\gamma}$. It follows that 
for a state $\psi\in {\cal H}_{geo,\gamma}$ the expectation value 
\be \label{4.32}
P_\psi((v,l),(v',l')):=
<\psi,\hat{P}((v,l),(v',l'))\psi>_{{\cal H}_{geo,\gamma}}
\ee
defines a positive definite, Hermitian matrix on ${\cal H}^1_\gamma$.
Of course, we are being here rather Cavalier concerning domain questions 
and self-adjoint extensions (one possible choice is the Friedrichs 
extension) but a more detailed analysis would go beyond the
exploratory purposes of this paper. (If $\Sigma$ is compact then 
$\gamma$ is a finite graph and all operators are bounded, although then 
we need to talk about boundary conditions). With these cautionary 
remarks out of the way we then can state the following theorem.
\begin{Theorem} \label{th4.2} ~~~~~~~~~~~~\\
i)\\
Suppose that the operators $\hat{P}((v,l),(v',l')),\hat{Q}((v,l),(v',l'))$
form an Abelian subalgebra of ${\cal L}({\cal H}_{geo,\gamma})$ and that
they are self-adjoint for each pair $(v,l),(v',l')$. 
Then there exists a unitary operator $\hat{U}$ on 
${\cal H}_{geo,\gamma}\otimes {\cal H}^1_\gamma$ such that
\ba \label{4.33} 
&&
\frac{1}{2}[<p,\hat{P} p>_{{\cal H}^1_\gamma}
+<q,\hat{Q} q>_{{\cal H}^1_\gamma}]
=<\hat{z}^\dagger,\hat{\omega} \hat{z}>_{{\cal H}^1_\gamma}
\nonumber\\
\hat{z}_l(v)&=&\frac{1}{\sqrt{2}}
\sum_{v',l'}[
\hat{a}((v,l),(v',l')) q_{l'}(v')-i\hat{b}((v,l),(v',l')) p_{l'}(v')]
\nonumber\\
\hat{a} &=& \hat{U} \hat{D} \nonumber\\
\hat{D}&=& 
\sqrt{
\sqrt{\hat{P}}^{-1}\;\;
\sqrt{
\sqrt{\hat{P}}\hat{Q}\sqrt{\hat{P}}
     }\;\;
\sqrt{\hat{P}}^{-1}
     }
\nonumber\\
\hat{b} &=& (\hat{a}^{-1})^T \nonumber\\
\hat{\omega} &=& (\hat{a}^{-1})^T\hat{Q}\hat{a}^{-1} 
\ea
where the square roots and inverses are with respect to 
${\cal H}_{geo,\gamma}\otimes {\cal H}^1_\gamma$ while transposition is 
with respect to ${\cal H}^1_\gamma$. 
Moreover, if we set 
\be \label{4.34}
\hat{\hat{c}}_l(v)=\frac{1}{\sqrt{2}}[
\sum_{v',l'}[
\hat{a}((v,l),(v',l')) \hat{q}_{l'}(v')-i\hat{b}((v,l),(v',l')) 
\hat{p}_{l'}(v')],
\ee
then $\hat{\hat{c}}_l(v),\hat{\hat{c}}^\dagger_l(v)$ satisfy the canonical 
commutation relations where the adjoint is with respect to 
${\cal H}_{geo,\gamma}\otimes{\cal H}_{matter,\gamma}$.\\
\\
ii)\\
Suppose that we are given real 
valued and symmetric operators $P,Q$ on ${\cal H}^1_\gamma$.
Then there exists a real valued, unitary operator $U$ on 
${\cal H}^1_\gamma$ such that
\ba \label{4.35} 
&&
\frac{1}{2}[<p,P p>_{{\cal H}^1_\gamma}
+<q, Q q>_{{\cal H}^1_\gamma}]
=<\bar{z},\omega z>_{{\cal H}^1_\gamma}
\nonumber\\
z&=&\frac{1}{\sqrt{2}}[a q-i b p]
\nonumber\\
a &=& U D
\nonumber\\
D &=& 
\sqrt{
\sqrt{P}^{-1}\;\; 
\sqrt{
\sqrt{P} Q\sqrt{P}
     }\;\;
\sqrt{P}^{-1}
     }
\nonumber\\
b &=& (a^{-1})^T \nonumber\\
\omega &=& (a^{-1})^T Q a^{-1} 
\ea
where the square roots, transposes and inverses are with respect to 
to ${\cal H}^1_\gamma$.
Moreover, if we set 
\be \label{4.36}
\hat{c}_l(v)=\frac{1}{\sqrt{2}}
\sum_{v',l'}[
a((v,l),(v',l')) \hat{q}_{l'}(v')-i b((v,l),(v',l')) 
\hat{p}_{l'}(v')]
\ee
then $\hat{c}_l(v),\hat{c}^\dagger_l(v)$ satisfy the canonical 
commutation relations where the adjoint is with respect to 
${\cal H}_{matter,\gamma}$.
\end{Theorem}
The proof is straightforward and is omitted. Notice that unitarity means that
$(\hat{U}^\dagger)^T=\hat{U}^{-1}$ where the dagger is with respect to 
${\cal H}_{geo,\gamma}$ while $U^T=U^{-1}$. The second assumption in i)
is not immediately satisfied for the operators $\hat{P},\hat{Q}$ in 
(\ref{4.18}) and (\ref{4.26}) as they stand because in the form displayed 
they are only symmetric, $(\hat{P}^\dagger)^T=\hat{P}$, and similar for 
$\hat{Q}$. However, since, with respect to the ${\cal H}^1_\gamma$ 
degrees of freedom they act on matter operators of the form $\hat{p}\otimes 
\hat{p}$
and since the $\hat{p}$ commute among each other, we may without loss of
generality assume that $\hat{P}^T=\hat{P}$ so that 
$\hat{P}=\hat{P}^\dagger$. The first assumption in i), however, is violated 
for the geometrical operators $\hat{P},\hat{Q}$ of (\ref{4.18}) and 
(\ref{4.36}), they do not commute on general states. On generic states, 
however, they do commute. This {\it non-commutative geometry} will lead to 
further quantum corrections for what follows, meaning that the operators
$\hat{\hat{c}},\hat{\hat{c}}^\dagger$ satisfy canonical commutation 
relations on generic states only but not exactly. We will not discuss these
effects in this paper and from now on assume that 
$\hat{P}((v,l),(v',l')),\hat{Q}((v,l),(v',l'))$ generate an Abelian operator
algebra on ${\cal H}_{geo,\gamma}$.\\
\\
With these cautionary remarks out of the way, theorem \ref{th4.2} suggests
to choose a different ordering for the operator (\ref{4.31}), namely
the {\it normal ordered form}
\be \label{4.37}
\hat{H}_\gamma=\sum_{(v,l),(v',l')}
\hat{\hat{c}}_l(v)^\dagger\hat{\omega}_{(v,l),(v',l')}\hat{\hat{c}}_{l'}(v').
\ee
When comparing (\ref{4.37}) with (\ref{4.31}) one finds out that 
they differ by a purely gravitational operator which is the quantization
of the usual, IR divergent, normal ordering constant in flat space. This
can be avoided as follows: The clean way to arrive at the form
(\ref{4.37}) from first principles is to write the classical expression 
\be \label{4.38}
H=\frac{1}{2}\int_\Sigma d^3x\int_\Sigma d^3y 
[P((x,l),(y,l') p_l(x) p_{l'}(y)+Q((x,l),(y,l') q_l(x) q_{l'}(y)]
\ee
whose quantization gives rise to (\ref{4.31}), {\it first} classically in 
the form 
\be \label{4.39}
H=\int_\Sigma d^3x\int_\Sigma d^3y \overline{c_l(x)}
\omega((x,l),(y,l')) c_{l'}(y)
\ee
and {\it then} to quantize it. However, in order to do that we would need,
for instance, the explicit expression of the functions $\omega((x,l),(y,l')$
in terms of the elementary gravitational degrees of freedom $A_a^j, E^a_j$
which is unknown. Thus, our procedure to first quantize (\ref{4.31}) and
then to normal order it should be considered as the ``poor man's way" of 
quantizing (\ref{4.39}) directly which would not lead to a {\it normal 
ordering operator}.
 
We have judiciously chosen the double hat notation for the operator 
$\hat{\hat{c}}$ in order to indicate that it involves {\it the 
anticipated mixture of gravitational and matter quantum operators}.\\
\\
{\bf 
We suggest that the operators $\hat{\hat{c}}_l(v)$ play the role 
of the fully geometry -- matter coupled system that is normally played 
by the matter annihilation operators on a given background geometry.}\\
\\
In order to justify this, we should now construct {\it coherent states}
of ${\cal H}_{geo,\gamma}\otimes{\cal H}_{matter,\gamma}$ 
that are eigenstates of $\hat{\hat{c}}$ and from them a {\it vacuum state}
and {\it $n$-particle states}. (Notice that such states are automatically
embedded in the full kinematical Hilbert space). We will choose the 
complexifier method \cite{Thiemann:2002vj} in order to do that.

The idea of a complexifier is to find an operator $\hat{C}$, which in our 
case will depend on both gravitational and matter degrees of freedom,
such that
\be \label{4.40}
e^{-\hat{C}/\hbar}\hat{q}e^{\hat{C}/\hbar}
=\sqrt{2}\hat{a}^{-1}\hat{\hat{c}}.
\ee
(As before, we are working on ${\cal H}_\gamma=
{\cal H}_{geo,\gamma}\otimes {\cal H}_{matter,\gamma}$ for each 
$\gamma$ separately).
Comparing with (\ref{4.33}) we find the unique solution 
\be \label{4.41}
\hat{C}=\hat{C}_{geo}+\frac{1}{2}\sum_{(v,l),(v',l')} 
\hat{p}_l(v)\hat{D}^{-2}((v,l),(v',l')) \hat{p}_{l'}(v')
\ee
where $\hat{C}_{geo}$ is a positive definite operator constructed from
the gravitational electrical degrees of freedom only according to the 
guidelines of \cite{Thiemann:2002vj} so that 
$[\hat{C}-\hat{C}_{geo},\hat{C}_{geo}]=0$. 

The complexifier coherent state
machinery can now be applied and we arrive at the following coherent 
states: Let $m$ be points in the full phase space of gravitational and 
matter degrees of freedom. Let $C$ be the classical limit of (\ref{4.14})
which we know in principle exactly in terms of $Q,P$ which were classically
given in terms of the gravitational three metric and partial derivative 
operators. Compute $z_g(m)=[e^{-i{\cal L}_{\chi_C}} A^E](m)$
and $z_m(m)=[e^{-i{\cal L}_{\chi_C}} q](m)$ where $A^E$ is the 
gravitational connection, $q=A^M$ or $q=\phi$ are the Maxwell connection and 
Klein Gordon scalar field respectively, $\chi_C$ is the Hamiltonian 
vector field of $C$ and $\cal L$ denotes the Lie derivative. Both 
functions are functions of both matter and geometry degrees of freedom.
Let $\delta_{h';\gamma}\otimes
\delta_{H';\gamma}$ be the $\delta$ distribution with respect to the uniform
measures of the $L_2$ spaces that define ${\cal H}_{kin,\gamma}$ and
denote by  
$h',H'$ the set of gravitational and matter (point) holonomies along the 
edges of $\gamma$. Then, for instance for Maxwell matter
\be \label{4.42}
\psi_{m;\gamma}:=(e^{-\hat{C}/\hbar} \;
[\delta_{h';\gamma}\otimes \delta_{H';\gamma}]
)_{h'\to h(z_E(m),H'\to H(z_M(m))}
\ee
define coherent states on ${\cal H}_{kin,\gamma}$ as shown in 
\cite{Thiemann:2002vj}.
Here $h(z_E(m))$ denotes the set of gravitational holonomies $h(A^E)$ 
along the edges of $\gamma$ where the real connection $A$ is replaced by 
the complex connection $z_E(m)$ (analytical extension) and similar
for $H(z_M(m))$. One of the nice features of the coherent states 
(\ref{4.42}) is the fact that they are {\it eigenstates} of the operators
$\hat{\hat{c}}_l(v)$ with eigenvalue $c_l(v)[m]$ as one can explicitly 
check (using the fact that $\hat{a}$ commutes with $\hat{D}$).

As an example, consider the case of photons propagating on fluctuations 
around flat (i.e. empty) space. Then we have 
$m^0_E:=(A_a^j,E^a_j)=(0,\delta^a_j)$ and $m^0_M:=(A_a,E^a)=(0,0)$ so that
for an edge $e\in E(\gamma)$ we have $H_e(z_M(m^0))=1$. 
We have $q_l(v)\equiv q_I(v)=\ln(H_{e_I(v)})/i$ so that 
$c_l(v)[m^0]=0$. Thus, the operators $\hat{c}_l(v)$ annihilate 
the {\it vacuum state} $\Omega_\gamma:=\psi_{m^0;\gamma}$ over $\gamma$.
Moreover, since 
$[\hat{\hat{c}}_l(v),\hat{\hat{c}}_{l'}(v')]=\hbar\delta_{v,v'} 
\delta_{l,l'}$
we are able, in principle, to construct a symmetric Fock space 
${\cal F}_{\gamma}({\cal H}^1_\gamma)$ where the $n-$particle states are 
defined by
\be \label{4.43}
\ket{f_1,\ldots ,f_n}:=
\hat{\hat{c}}^\dagger(F_1)\ldots\hat{\hat{c}}^\dagger(F_n)\Omega_\gamma
\mbox{ where }
\hat{\hat{c}}^\dagger(F):=\sum_{v,l} F_l(v)\hat{\hat{c}}^\dagger_l(v).
\ee
A precise map between (\ref{4.43}) and the usual photon states on 
flat Minkowski space can be given for the fundamental states (\ref{4.43})
as well but we postpone this to the next section.

Notice that due to commutativity of different matter types we can 
add the operators $\hat{C}$ in (\ref{4.41}) for different matter types in
order to arrive at simultaneous coherent and Fock states for all matter 
types!

\subsection{Approximate Fock States and Semiclassical States}
\label{s4.4}

It is clear that the program sketched in section \ref{s4.3} cannot be 
carried out with present mathematical technology because we are not really
able to construct the operators $\hat{a},\hat{b},\hat{\omega}$ which 
require precise knowledge of the spectrum of these operators on 
${\cal H}_{geo,\gamma}\otimes {\cal H}^1_\gamma$. Thus, in order to 
to proceed we have to do something {\it much more moderate}.
As a first approximation we consider states which do not mix matter and 
geometry degrees of freedom in the way (\ref{4.42}) did but rather 
will look for Fock states of the form $\psi_{m_E;\gamma}\otimes 
\psi_{\text{matter};\gamma}$ where $\psi_{m_E;\gamma}$ is a gravitational 
coherent state peaked at the point $m_E\in {\cal M}_E$ 
in the purely gravitational phase space, for instance the ones constructed in 
\cite{Thiemann:2000bw,Thiemann:2000ca,Thiemann:2000bx}. 
These states are generated by the piece $\hat{C}_{geo}$ of 
the complexifier of (\ref{4.41}) by applying its exponential to the 
$\delta$ distribution on the gravitational Hilbert space alone. Let 
$\{T_n\}$ be 
a complete orthonormal basis of states in ${\cal H}_{\text{matter},\gamma}$.
Using the overcompleteness of the just mentioned coherent states,
we can write the matrix elements of $\hat{H}_\gamma$, given in the normal 
ordered form of (\ref{4.37}) as (interchange of summation and 
integration must be justified by closer analysis)
\ba \label{4.44}
&& <\psi_{m_E;\gamma}\otimes \psi_{\text{matter}},
\hat{H}_\gamma \psi_{m_E;\gamma}\otimes \psi'_{\text{matter}}>
\nonumber\\
&=& \sum_{(v,l),(v',l')} 
\int_{{\cal M}^\gamma_E} d\nu_\gamma(m_E)
\int_{{\cal M}^\gamma_E} d\nu_\gamma(m'_E)\sum_{n,n'}
<\hat{\hat{c}}_l(v) \psi_{m_E;\gamma}\otimes \psi_{\text{matter}},
\psi_{m_E;\gamma}\otimes T_n>\;\times \nonumber\\
&& \times
<\psi_{m_E;\gamma}\otimes T_n,\hat{\omega}_{(v,l),(v',l')}
\psi_{m'_E;\gamma}\otimes T_{n'}>\;
<\psi_{m'_E;\gamma}\otimes T_{n'},\hat{\hat{c}}_{l'}(v')
\psi_{m_E;\gamma}\otimes \psi'_{\text{matter}}>.
\ea
Here $\nu_\gamma$ is the Hall measure \cite{Hall1994} generalized to graphs 
in \cite{Ashtekar:1996nx} and ${\cal M}^\gamma_g$ is ${\cal M}$ restricted to the 
graph $\gamma$ as defined in \cite{Thiemann:2000bv}.

Let us introduce the real valued operators on ${\cal H}^1_\gamma$ defined by
\be \label{4.45}
a^{m_E}((v,l),(v',l'):=<\psi_{m_E},\hat{a}((v,l),(v',l'))\psi_{m_E}>_{{\cal 
H}^E_\gamma}
\ee
and similar for $\hat{\omega},\hat{b}$. Consider also the operators on
${\cal H}_{\text{matter},\gamma}$ defined by
\be \label{4.46}
\hat{c}^{m_E}_l(v):=\frac{1}{\sqrt{2}}\sum_{v',l'}
[a^{m_E}((v,l),(v',l')) \hat{q}_{l'}(v')-i 
b^{m_E}((v,l),(v',l')) \hat{p}_{l'}(v')].
\ee
The coherent states $\psi_{m_E;\gamma}$ are sharply peaked in ${\cal 
M}^\gamma$
which implies that up to $\hbar$ corrections 
\be \label{4.47}
<\psi_{m_E;\gamma}\otimes \psi_{\text{matter}},
\hat{\hat{c}}
\psi_{m'_E;\gamma}\otimes \psi'_{\text{matter}}>
=\delta_{\nu_\gamma}(m_E,m'_E)
<\psi_{\text{matter}},\hat{c}^{m_E}\psi'_{\text{matter}}>
\ee
and similar for $\hat{\omega}$. We conclude that up to $\hbar$ 
corrections 
\ba \label{4.48}
&& <\psi_{m_E;\gamma}\otimes \psi_{\text{matter}},
\hat{H}_\gamma \psi_{m_E;\gamma}\otimes \psi'_{\text{matter}}>
\nonumber\\
&=& \sum_{(v,l),(v',l')} 
<\hat{c}^{m_E}_l(v) \psi_{\text{matter}},\omega^{m_E}_{(v,l),(v',l')}
\hat{c}^{m_E}_{l'}(v')\psi'_{\text{matter}}>.
\ea
Now, again using that the $\psi_{m_E;\gamma}$ have very strong 
semiclassical
properties, it is possible to show that, up to $\hbar$ corrections, the
operators $a^{m_E},b^{m_E},\omega^{m_E}$ can be computed by {\it first} 
calculating 
the expectation values of the operators $\hat{P},\hat{Q}$ on 
${\cal H}^E_\gamma\otimes{\cal H}^1_\gamma$, which we know explicitly, to 
arrive at operators $P^{m_E},Q^{m_E}$ on ${\cal H}^1_\gamma$ and {\it 
then}
to plug those into the formulas (\ref{4.35}). Thus, we have arrived at a 
``poor man's version" of an annihilation and creation operator 
decomposition of $\hat{H}_\gamma$ which approximates the exact version
but which still takes the fluctuating nature of the quantum geometry into
account through the uncertainties encoded into the states 
$\psi_{m_E;\gamma}$.

It is now clear how we arrive at approximate Fock states. Instead of
(\ref{4.41}) we consider the operator on ${\cal H}_{\text{matter};\gamma}$
defined by
\be \label{4.49}
\hat{C}^{m_E}=\frac{1}{2}\sum_{(v,l),(v',l')} 
\hat{p}_l(v) (D^{m_E})^{-2}((v,l),(v',l')) \hat{p}_{l'}(v').
\ee
This complexifier now generates coherent states on ${\cal H}_{\text{matter},\gamma}$ 
in analogy to (\ref{4.42}), e.g. for Maxwell matter, by
\be \label{4.50}
\psi^{m_E}_{m_M;\gamma}:=(e^{-\hat{C}^{m_g}/\hbar} 
\delta_{H';\gamma})_{H'\to H(z^{m_E}(m_M))},
\ee
where $m_M$ is a point in the matter phase space, 
$z^{m_E}_M(m_M)=(e^{-i{\cal L}_{\chi_{C^{m_E}}}} q)(m_M)$ and 
$C^{m_E}(m_M)$ is 
the classical limit of $\hat{C}^{m_E}$ on the matter phase space.
Choosing the points $m^0_E,m^0_M$ appropriate for vacuum will now 
produce a vacuum state $\Omega^0_\gamma:=\psi^{m^0_E}_{m^0_M;\gamma}$
for the operators $\hat{c}^{m_E^0}$ which by construction (theorem 
\ref{th4.2}) satisfy canonical commutation relations {\it exactly}.
Summarizing, with 
\be \label{4.51}
\psi_{m_E;\gamma}=(e^{-\hat{C}_{geo}/\hbar} \delta_{h'})_{h'\mapsto 
h(z'_E(m_E))},\;\;
z'_E(m_E)=e^{-i{\cal L}_{\chi_{C_{geo}}}} A^E
\ee
we arrive at {\it approximate} $n-$particle states
\be \label{4.52}
\ket{F_1,\ldots,F_n}'=\psi_{m_E,\gamma}\otimes 
\hat{c}^\dagger(F_1)\ldots\hat{c}^\dagger(F_n)\Omega^0_\gamma,
\ee
where $\hat{c}^\dagger(F)=\sum_{v,l} F_l(v)\hat{c}^\dagger_l(v)$
and the associated Fock space.

Let us conclude this section with some remarks:\\
1)\\
For the lepton sector things are much easier because the operators 
$\hat{\theta}^\mu_A(x),(\hat{\theta}^\mu_A(x))^\dagger$ already satisfy 
canonical commutation relations and all momenta are ordered to the 
left in $\hat{H}^D_\gamma$. A suitable vacuum state annihilated 
by all the $\hat{\theta}^\mu_A(v)$ is given (up to normalization) by 
\be \label{4.53}
\Omega^D_\gamma(\theta)=\prod_{v,A,\mu} \theta^\mu_A(v).
\ee
2)\\
In order to relate, say the states (\ref{4.52}) to the usual Fock states 
on Minkowski space with Minkowski vacuum $\Omega_F$ and smeared creation
operators $\hat{c}_F(f)=\int d^3x f_L(x) (\hat{c}^L_F)^\dagger(x)$ 
with different labels $L$ we need 
the particulars of the expectation values of the gravitational operators.
Basically, the discrete sum involved in the definition of $\hat{c}(F)$
is a Riemann sum approximation to the integral involved in the definition 
of $\hat{c}_F(f)$ which becomes exact in the limit that the lattice spacing
$\epsilon$ vanishes where $F^f_l(v):=\epsilon^n X_l^L(v) f_L(v)$ for some 
power of $\epsilon$ and some matrices $X_l^L$ which depend on the choice 
of the gravitational coherent states. One can then establish a map 
between the usual Fock states and our graph dependent ones by
\be
\hat{c}_F^\dagger(f_1)\ldots\hat{c}_F^\dagger(f_n)\Omega_F
\mapsto 
\hat{c}^\dagger(F^{f_1})\ldots\hat{c}^\dagger(F^{f_n})\Omega_\gamma
\ee
which becomes an isometry in the limit $\epsilon\to 0$!
In more detail, let us consider the simple example of a regular cubic 
lattice in $\Rl^3$. Discarding fluctuation effects from the gravitational 
field we arrive at the dimensionfree graph annihilation operators
for Maxwell theory given by
\be 
\hat{c}^I_\gamma(v)=\frac{1}{\sqrt{2\alpha}}[
\root[4]\of{-\Delta_\gamma} \hat{A}_I
-i\root[4]\of{-\Delta_\gamma}^{-1} \hat{E}^I](v),
\ee
where $\Delta_\gamma=\delta^{IJ} \partial^-_I\partial^+_J$.
On the other hand, for the usual Fock representation we get the 
annihilators of dimension cm$^{-3/2}$ given by
\be 
\hat{c}^a_F(x)=\frac{1}{\sqrt{2\alpha}}[
\root[4]\of{-\Delta} \hat{A}_a
-i\root[4]\of{-\Delta}^{-1} \hat{E}^a](x).
\ee
Using dimensionfree transversal fields $F^I(v),\;\partial_I^- F^I=0$
on the polymer side and transversal fields 
$f^a(v),\;\partial_a f^I=0$ of dimension cm$^{-3/2}$ we arrive at
smeared, dimensionfree creation operators of the form
\be \label{4.56}
\hat{c}^\dagger_\gamma(F)=\sum_v F^I(v)\hat{c}_\gamma^I(v)
\mbox{ and } 
\hat{c}^\dagger_F(f)=\int d^3x\; f^a(x)\hat{c}_F^a(x).
\ee
Using $E^I(v)\approx \epsilon^2\delta^I_a E^a_x,\;
A_I(v)\approx \epsilon \delta_I^a A_a(x),\;\Delta_\gamma(v)
\approx \epsilon^2 \Delta(v)$ we see that 
$\hat{c}_\gamma^I(v)\approx \epsilon^{3/2} \delta_a^I \hat{c}_F^a(v)$
so that our desired map is given by
\be \label{4.57}
(F^f)^I(v)=\delta^I_a(v) \epsilon^{3/2} f^a(v),
\ee
where the two factors of $\epsilon^{3/2}$ combine to the Riemann sum
approximation $\epsilon^3$ of the Lebesgue measure $d^3x$.

Finally we mention that the way, approximate $n$-particle states were
obtained above, bears some similarity 
to the treatments of gravitons in \cite{IR1,IR2}. In both cases, a
semiclassical state for the gravitational field is used to obtain a
classical background geometry. In \cite{IR1,IR2} a weave state is used 
for this purpose, here we have employed the coherent state
$\psi_{m_E}$. In other respects, the treatments differ, however. To
define the notion of gravitons, a split of the gravitational field in a
dynamical and a background part is necessary whereas nothing of this
sort is required for the treatment of matter fields. 
\section{Towards Dispersion Relations}
\label{se5}
Dispersion relations are the 
relations between the frequency $\omega$ and the wave vector $\evec{k}$ 
of waves of a field of some sort, traveling in vacuum or through some
medium. In quantum mechanical systems, the dispersion relation 
is the the relation between the momentum and the energy of particles. 
The form of the dispersion relations appearing in fundamental physics
is dictated by Lorenz invariance. Since this invariance is likely to be broken
in quantum gravity, modification of dispersion relations is
conjectured to be an observable effect of quantum gravity.  
In this section we would like to explain why QGR
indeed leads to modified dispersion relations, and how one might 
proceed in a calculation of these modifications.

There are at least two mechanisms by which modified dispersion
relations arise in the context of QGR, and it is
important to keep them apart. Let us start to discuss the first one by 
considering an analogous effect in another branch of physics: 

A prime example coming to mind when thinking about modified
dispersion relations is the propagation of light in materials. The
mechanism which causes these modification is roughly as follows:  
The electromagnetic field of the in-falling wave acts on the charges in the
material, they are accelerated and in turn create electromagnetic
fields. These fields interfere with the in-falling ones, the net effect 
of this is a wave with modified phase and therefore, a phase velocity
differing from the one in vacuum.  
The precise relation between the force acting on the charges and the
fields induced by them depends on the properties of the material and
also on the frequency of the wave, and thus gives rise to a frequency 
dependent phase velocity and, hence, a nontrivial dispersion relation. 
Under some simplifying assumptions, this relation looks as follows: 
\begin{equation*}
\omega(\betr{k})=\betr{k}\left(1-
\frac{\kappa}{\omega_0^2-\omega^2(\betr{k})+i\rho\omega(\betr{k})}\right)
\end{equation*}
where $\omega_0$, $\kappa$ and $\rho$ are properties of the material. 
As is to be expected, if the energy of the in-falling wave is very low compared
with the binding energies ($\sim \omega_0$) of the charges, the
frequency dependence of the phase velocity will also be very small. 

In QGR, modified dispersion relations can be expected 
from the interplay between matter and quantum gravity by an analogous 
mechanism: The propagating matter wave causes changes in the local
geometry, which in turn affect the propagation of the wave. Again, if
the energy of the wave is very small, so will be the modification of
the dispersion relation as compared to the standard one. 

In order to honestly account for this back-reaction mechanism one would 
have to do a first principle calculation that involves solving the 
combined matter -- geometry Hamiltonian constraint. Technically, we are not
yet in the position to do that. However, as a first approximation we 
{\it can} take care of the reaction of the geometry to the matter
fields by using coherent states $\Psi_{m_E}$, $\Psi_{\text{matter}}$
in the constructions of section \ref{se4} which are peaked at a
classical configuration which is a solution of the field equations of
the combined gravity-matter system. 
This should be seen
in analogy to our remarks in section 2 where QED corrections are computed 
with coherent states for {\it free} Maxwell theory instead of the full QED
Hamiltonian which neglects the back-reaction from the fermions.

There is, however, a second source of modifications to the dispersion
relations: The inherent discreteness of geometry found in QGR. 
This effect has nothing to do with back-reaction of
the geometry on the matter and it is the contribution of this effect 
to the dispersion relations that can be studied with more confidence.
This is what we will discuss in the rest of this
section and in our companion paper \cite{ST02}. 

Let us again start by briefly reviewing an analogous phenomenon from a
different branch of physics, the propagation of lattice vibrations 
(``sound'') in crystals. As an example, consider an extremely simple model, a one
dimensional chain of atoms. We assume that all atoms have the same
mass $m$ and that each of them acts on its two neighbors with an attractive
force proportional to the mutual distance. 
If we denote by $\epsilon$ the interatomic distance in the equilibrium 
situation, by $q(z)$ the displacement of atom $z$ from its
equilibrium position $\epsilon z$ and set $p(z)=\dot{q}(z)$, the
Hamiltonian for the system reads
\begin{equation}
\label{eq5.1}
  H=\frac{1}{2}\sum_{z\in \Z}\frac{1}{m}p^2(z)+ K\left(q(z+1)-q(z)\right)^2.
\end{equation}
The corresponding equations of motion are simple, a complete set of
solutions is given by 
\begin{equation}
\label{eq5.2}
q(t,z)= \exp i\left(\epsilon z k -\omega(k)t\right),\quad \text{ with }\quad
\omega^2(k)=\frac{2K}{m}\left(1-\cos k\epsilon\right).  
\end{equation}
As the solutions are straightforward analogs of plane waves in the
continuum, $\omega^2(k)$ is readily interpreted as the dispersion
relation for the system. We see that it contains the ``linear'' term
proportional to $k^2$ expected for sound waves in the continuum, as well as
higher order corrections due to the discreteness of the lattice. 

Let us reconsider \eqref{eq5.1}: The fact that the $q(z)$ are
displacements of atoms is not explicitly visible. $H$ could as well be 
the Hamiltonian of a field $q$ with a certain form of potential, propagating
on a {\it regular lattice}. Having made that observation, we are already
very close to the model just described.
Upon
choosing a semiclassical state $\Psi$ for the
gravitational field,
the bosonic Hamiltonians of section \ref{se4} are of the form 
\begin{equation}
\label{eq5.3}
\hat{H}^\Psi_\gamma=\frac{1}{2}\sum_{v,v',l,l'} 
\hat{p}_l(v)P_\Psi((v,l),(v',l')) \hat{p}_{l'}(v') 
+\hat{q}_l(v)Q_\Psi((v,l),(v',l')) \hat{q}_{l'}(v'), 
\end{equation}
where the expectation values
\begin{equation*}
P_\Psi((v,l),(v',l'))=\expec{\hat{P}((v,l),(v',l'))}_{\Psi}\qquad 
Q_\Psi((v,l),(v',l'))=\expec{\hat{Q}((v,l),(v',l'))}_{\Psi}. 
\end{equation*}
contain an imprint of the fluctuations of the gravitational field.
$\Psi$ can in principle be taken to be a coherent state for the 
gravitational field peaked at an arbitrary 
point of the classical phase space. However, since we are interested
in dispersion relations, a notion that by definition describes the
propagation of fields in flat space, we will restrict considerations
to the case of GCS approximating flat Euclidean space (denoted by
$\Psi_{\text{flat}}$ in the following).\footnote{Also, when
  considering application to situations such as the $\gamma$-ray 
burst effect, the curvature radius is always huge compared to
Planck length and does therefore not lead to any new quantum effects 
but just to classical redshifts which can easily be accounted for.}   

There are however two fundamental differences between \eqref{eq5.1}
and \eqref{eq5.3}, and we will discuss them in succession. 
The first difference is that
\eqref{eq5.3} is a Hamiltonian for a \textit{quantum} field whereas the
former is purely classical. 
Note however, that the Hamiltonians of section \ref{se4} are normal ordered. 
Thus, the expectation value of these Hamiltonians in a
coherent state peaked at a specific classical field configuration
(defined in sections \ref{s4.3}, \ref{s4.4}) will 
yield \textit{precisely} its classical value. Moreover, expectation
values for the matter quantum fields will exactly equal the the
corresponding classical values at all times. 
Therefore, in discussing the dispersion relations, we will assume the
matter quantum fields to be in a coherent state and can  
effectively work with the classical fields $p,q$.
This will be a very good 
approximation in processes such as light propagation due to decoherence 
of matter quantum effects in large ensembles of photons and because
higher loop corrections of QED are not Poincar\'e violating (which is 
what we are interested in here). 
 
The second and more important difference between  \eqref{eq5.1}
and \eqref{eq5.3} lies in the following: In \eqref{eq5.1}, the
coefficients of the fields do not depend on the vertex. This is the
reason why one can explicitly calculate solutions to the equations of
motion. In contrast to that, $P((v,l),(\,\cdot\,,l'))$ will in general 
depend on $v$, even if the state $\Psi_{\text{flat}}$
employed to compute the gravity expectation values is a good 
semiclassical state. 
As a result, the field equations will be 
complicated and, most important for us, not have ``plane wave''
solutions
\begin{equation}
\label{eq5.5}
q_{\evec{k}}(t,v)=\exp i(\evec{k}\evec{x}(v)-\omega t)  
\end{equation}
any more.   
Hence if we would Fourier decompose solutions of the field equations 
with respect to \eqref{eq5.5}, the support of the resulting functions 
will not be confined by a
dispersion relation to some line in the $\omega$-$\betr{k}$ plane,
anymore. 
 
\begin{figure}
\centerline{\epsfig{file=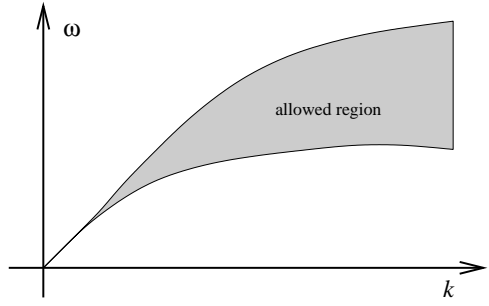, height=4cm}}
\caption{\label{fi5.1}Fourier transform of EOM. The support of solutions has to lie
in the shaded region.}
\end{figure}

\begin{figure}
\centerline{\epsfig{file=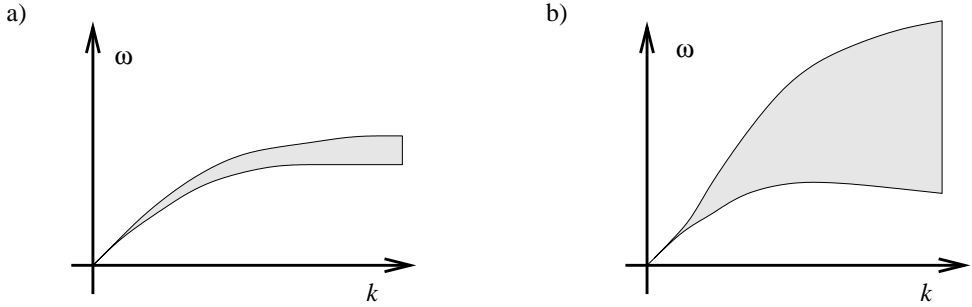, height=4cm}}
\caption{\label{fi5.2}Can higher order corrections to the dispersion relation be
  given? a) Yes, approximately. b) No, there is no meaningful notion of
  dispersion relation beyond linear order}
\end{figure}

However, for a good semiclassical state, symmetry, which is
absent due to the vertex dependence of the coefficients, will be
approximately restored on a large length scale.  For example, if 
the vertex dependent coefficients would be averaged over large enough
regions of $\Sigma$ the average would be independent of the specific
choice of the region. Therefore, for long wavelength, 
plane waves \eqref{eq5.5} should at least be approximate
solutions to the field equations. The following scenario is conceivable: 
Although there is no exact dispersion relation, the support of the
Fourier transform of a solution might be confined to some region in the
$\omega$-$\betr{k}$ plane, or the Fourier transform has at least to be 
peaked there. This region should get more and more narrow for 
longer wavelength, 
leading to an ordinary dispersion relation in the limit (see figure 
\ref{fi5.1}).
We have to note, however that even if this is true, there is no
guarantee that a dispersion relation with corrections to the linear term 
makes sense as an approximate description for long wavelength.  
We tried to visualize this in figure \ref{fi5.2}.  
So, to conclude, it is very plausible that a nonlinear dispersion relation
will turn out to be a good approximate description of the physical
contents of \eqref{eq5.3} for long wavelength in this sense.  
But issues such as the one depicted in figure \ref{fi5.2} definitely
merit further studies. 

Let us now turn to the practical question of how a nonlinear
dispersion relation can actually be computed from \eqref{eq5.3}. A
simple model that displays some of the complications due to the vertex 
dependence of the coefficients in \eqref{eq5.3}, but can nevertheless
be treated with analytical methods, can be obtained from \eqref{eq5.1}  
in the following way: Upon setting $m=1$ and $K=l^{-2}$, the
Hamiltonian \eqref{eq5.1} can be interpreted as that of a scalar field 
propagating on a one dimensional lattice with lattice spacing $l$. Now 
we partly remove the assumption of a constant lattice spacing by
replacing $l$ by $l_z$ -- the distance between the lattice point labelled
by $z$ and the one labelled by $z+1$ --  
but still assume periodicity on a large scale, 
\begin{equation*}
  l_z=l_{N+z}\quad\text{for all }\quad z\in \Z
\end{equation*}
for some $N\in\N$. It turns out that the dispersion relation for this
system has several branches and the small $k$ behavior of the
acoustic branch is given by
\begin{equation}
\label{disp}
  \omega_{\text{ac}}^2(k)=
  \frac{\aver{l}{}^2}{\aver{l^2}{}}\betr{k}^2
  +\left(\frac{1}{L^2}\frac{\aver{l}{}^6}{\aver{l^2}{}^3}
    \sum_{i<j}c_{ij}l_i^2l_j^2
    -\frac{L^2}{12}\frac{\aver{l}{}^2}{\aver{l^2}{}}
  \right)\betr{k}^4+O\left(\betr{k}^6\right). 
\end{equation}
Here, 
\begin{equation*}
  L=\sum_{n=0}^{N-1}l_n,\qquad c_{ij}=(j-i)[N-(j-i)],\quad\text { and }
\aver{l}{}=\frac{1}{N}\sum_{n=0}^{N-1}l_n\quad\text{ etc.}
\end{equation*}
As we we have indicated by means of notation, it is instructive to
view the $l_i, i=0,\ldots,N-1$ as independent random variables. The
moments of the corresponding distribution determine the dispersion
relation \eqref{disp}. In each order in $k$, corrections are present as
compared to the case of constant lattice spacing. Note in particular
that the phase velocity $\lim_{k\rightarrow 0} \omega(k)/k$ can be
smaller than one. 

It is remarkable how subtle the dependence of the dispersion relation
on the distribution of the lengths is already in this simple model. 
Although we do not have a proof, it is plausible that qualitatively, 
the above formula extends to higher dimensions, where analytical
proofs get much harder. For more information about the model as well
as a proof of \eqref{disp} we refer the reader to \cite{thesis} or
\cite{ST03}, for a beautiful numerical analysis of similar models in
two dimensions to \cite{Tanguy:2002}.\\    

The discussion of the one dimensional model given above shows how
complicated an exact analysis of the equations of motion is already in 
simple cases. Since the models 
\begin{equation}
\label{eq5.4}
H^{\Psi_{\text{flat}}}=\frac{1}{2}\sum_{v,v',l,l'} 
p_l(v) P_{\Psi_{\text{flat}}}((v,l),(v',l'))p_{l'}(v') 
+q_l(v) Q_{\Psi_{\text{flat}}}(v,l),(v',l')) q_{l'}(v'), 
\end{equation}
that one obtains from QGR are more complicated, it is useful to
explore a less precise but easier route towards dispersion relations. 
The idea which we would like to advocate  
is to replace \eqref{eq5.4} by a simpler Hamiltonian which
\begin{itemize}
\item is a good approximation of \eqref{eq5.4} for slowly varying
  $q$ and $p$ and  
\item is simple enough such that the EOM can be solved exactly. 
\end{itemize}
This idea underlies also the works \cite{Gambini:1998it} and 
\cite{Alfaro:1999wd} and, at a rather
simple level, is the basis for the recovery of continuum elasticity
theory from the atomic description in solid state physics 
(see for example \cite{Ashcroft:1976}). 

We will now propose a replacement for \eqref{eq5.4} fulfilling the
above requirements. In the long wavelength
regime, we can revert to the continuum picture, i.e. replace the
lattice fields 
$q_l(v),p_l(v)$ by $q_l(\evec{x}(v)),p_l(\evec{x}(v))$, where now
$q_l,p_l:\Sigma\rightarrow \C$.  
 
We should however take care that information about the lattice is at
least partially encoded in the continuum theory. To this end, we
Taylor expand the (now continuum) fields in the Hamiltonian up to a
certain order. For example, we would make the replacement
\begin{multline*}
q_l(\evec{x}(v)) Q_\Psi((v,l),(v',l')) q_{l'}(\evec{x}(v'))\\
\vla q_l(\evec{x}(v)) Q_\Psi((v,l),(v',l')) 
\Big[ b^i(v,v')\partial_iq_{l'}(\evec{x}(v))+
\frac{1}{2!}b^i(v,v')b^j(v,v')
\partial_i\partial_jq_{l'}(\evec{x}(v))+\ldots\Big]
\end{multline*}
where $b^i(v,v')\doteq \evec{x}(v')-\evec{x}(v)$. Even if we terminate 
the Taylor expansion after a few terms, the resulting Hamiltonian will 
be an excellent approximation to the original one, provided $p$ and
$q$ change only very little from vertex to vertex, our standing
assumption in the whole procedure. 

Now we will eliminate the spacial dependence of the coefficients,
which was the main difficulty in dealing with the original Hamiltonian  
\eqref{eq5.4}, by replacing them by their averages over all vertices of 
the graph. This can be justified as follows: 
If the fields $p,q$ are varying considerably 
only on length scales much larger then some macroscopic scale $L$, 
it is a very good approximation to replace the vertex dependent 
coefficients in the Hamiltonian by their averages over the vertices 
in regions of dimension $L^3$. On the other hand, as we have said
before, a good semiclassical
$\Psi_{\text{flat}}$ state will insure that the system described by 
\eqref{eq5.4} has the symmetries of flat space at least at large
distances. One way to state this more precisely is that 
the average of the coefficients appearing 
in \eqref{eq5.4} over vertices in regions with characteristic
dimension $L^3$ or larger, is \textit{independent} of the region to a
good approximation. Therefore we can indeed replace the vertex
dependent coefficients by their averages over all vertices. 

Let us again give an example for a typical term in the Hamiltonian: 
\begin{equation*}
\sum_{v'} Q((v,l),(v',l'))
b^{a_1}(v,v')\ldots b^{a_n}(v,v')=:F^{a_1\ldots a_n}(v,l,l') 
\vla \aver{F^{a_1\ldots a_n}(\cdot,l,l')}{}
\end{equation*}
where we have introduced the graph average
\begin{equation}
\label{av}
  \aver{F^{a_1\ldots a_n}(\cdot,l,l')}{}\doteq \frac{1}{N}\sum_v 
   F^{a_1\ldots a_n}(v,l,l'),
\end{equation}
$N$ being the number of vertices of the graph $\Psi_{\text{flat}}$ is
based on. In case we are dealing with an infinite number of vertices, 
the definition \eqref{av} has to be replaced by the limit of averages
over finite but larger and larger numbers of points.   

Finally we can replace the sum over vertices of \eqref{eq5.5} by 
an integral. We thus end up with a Hamiltonian for a continuum field
theory on $\Sigma$ and the coefficients of the fields being
constant. Therefore the equations of motion of the theory admit plane
waves as solutions, and their dispersion relation can be computed and
discussed. This dispersion relation should describe the
physical content of \eqref{eq5.4} for low energies (large wavelength). 

To justify our procedure, let us point out again, that it uses both
assumptions (large wavelength -- homogeneity and isotropy of the state on large
scales) that seem essential from physical considerations, to recover a 
dispersion relation from \eqref{eq5.4}, enter in a transparent way. 
Also, if we apply the procedure outlined above 
to the simple regular lattice system
\eqref{eq5.1}, we recover, order by order, the nonlinear dispersion
relation \eqref{eq5.2}. Thus, at least in this example, the simplified 
continuum theory still captures the information about the lattice to
any desired order of accuracy.    

In the companion paper \cite{ST02} we will elaborate on the procedure
described above and 
apply it to derive approximate dispersion relations from the
Hamiltonians constructed in this paper, evaluated in the gauge theory
coherent states of \cite{Thiemann:2000bw}. 
\section{Summary and Outlook}
\label{se6}
The goal of the present work was to begin investigations of the structure and
semiclassical limit of the theory obtained by coupling matter fields 
to QGR. A basic assumption that we made was
that the complicated dynamics
of a full theory could be approximated by 
treating the matter parts in the Hamilton constraint of the full 
theory as Hamiltonians generating the
matter dynamics and by the use of semiclassical states in the
gravitational sector. 

Using this assumption we obtained the following results:
\begin{enumerate}
\item We have proposed quantum theories of scalar, electromagnetic and 
  fermionic fields coupled to QGR. The dynamics of these theories is
  generated by a Hamiltonian in the same way as in ordinary QFT. Consequently
  we were able to identify approximate $n$-particle states which
  correspond to the usual Fock states for matter fields propagating  
  on classical geometries. 
  In other respects, the theories are very different from ordinary
  QFT, thus reflecting basic properties of QGR:
\begin{itemize}
\item The basic excitations of the gravitational field in QGR are
  concentrated on graphs. The requirement of diffeomorphism 
  invariance forces the
  matter degrees of freedom to be confined to the same graph as the 
  gravitational field. The matter fields are therefore 
  bound to become quantum fields propagating on a discrete 
  structure. 
\item In ordinary QFT, the background metric enters the definition of
  the ground state and the commutation relations of the fields. 
  In QGR on the other hand, the geometry is a dynamical variable,
  represented by suitable operators.  
  A QFT coupled to QGR therefore has to contain these operators
  in its very definition.   
  This is reflected in the theories of section \ref{se4} by the fact that their
  annihilation and creation operators act on both, the one particle
  Hilbert space of the matter fields \textit{and} the Hilbert space  
  of the geometry. 
\end{itemize}
We also showed how a ``QFT on curved space-time limit'' can be
obtained from this theory, using a semiclassical state of the
gravitational field.  
\item We have discussed how modified dispersion relations for the
  matter fields arise in the context of QGR and motivated a method for 
  computing them from the (partial) expectation values of the 
  quantum matter Hamiltonians in a semiclassical state.
\end{enumerate}
Certainly, the present work can only be regarded as a first step
towards a better understanding of the interaction of matter and
quantum gravity. 
In future work, the assumptions that have been used 
should be removed, or their validity confirmed. 

On the other hand, application of the results of the present work can
be envisioned. For example, it will be very interesting to see, whether 
the methods used in the present work can also be applied to
investigate how gravitons arise in the semiclassical regime of QGR. 
This will be the topic of \cite{QFTQGR}. 
As another application, the companion paper \cite{ST02} contains 
a calculation of corrections to the standard dispersion relations for
the scalar and the electromagnetic field due to QGR.

To summarize, the interaction of quantum matter and quantum gravity is 
a fascinating but, alas, very complicated topic, of which a good
understanding still has to be gained. 
We hope that the present work illuminates the difficulties encountered 
in this endeavor and also contains some first, albeit small, steps 
towards its completion.
\\
\\
\\ 
{\bfseries \Large Acknowledgements}
\\
\\
It is a pleasure for us to thank Abhay Ashtekar, Luca Bombelli, 
Arundhati Dasgupta, Rodolfo Gambini, Jurek Lewandowski, Fotini
Markopoulou Kalamara, Hugo Morales-Tecotl, Jorge Pullin,
and Oliver Winkler for numerous valuable discussions.  
We also thank the Center for Gravitational Physics of The
Pennsylvania State University, where part of this work was completed, for 
warm hospitality. T.T. was supported in part by NSF grant PHY 0090091 to The
Pennsylvania State University.

H.S. also gratefully acknowledges the splendid hospitality at the 
Universidad Autonoma Metropolitana Iztapalapa, Mexico City, and at 
the University of Mississippi, as well as the financial support by the
Studienstiftung des Deutschen Volkes.  
\begin{appendix}
\section*{Appendix: Kinematical vs. Dynamical Coherent States: 
A Simple Example} 
\setcounter{section}{1}
\label{ap1}
In systems with constraints linear in the basic variables, the
expectation values of Dirac observables 
in a coherent state in the kinematical Hilbert space equal those in
a dynamical coherent state, provided that both states are chosen to be peaked
around the same point in the constrained phase space. This does not
hold true anymore for systems with nonlinear constraints. One expects, 
however that the discrepancies between the expectation values on the
kinematical and on the dynamical level will at least be small. 
In this appendix, we demonstrate that this is true for a simple
quantum mechanical model system with a nonlinear constraint: A system 
of two coupled harmonic oscillators. 

Let the Hamiltonian of the harmonic oscillator be given as 
\begin{equation*}
H=\frac{1}{2}\left(\frac{p^2}{m}+ m\omega^2q^2\right). 
\end{equation*}
It is well known that it can be quantized in terms of annihilation and 
creation operators
$\widehat{a},\widehat{a}^\dagger$,
$\,\comm{\widehat{a}}{\widehat{a}^\dagger}= 1$ on the Fock space
$\hilb{H}$ over $\C$. $\widehat{a}$ is the quantization of the
classical quantity
\begin{equation*}
  z=\sqrt{\frac{m\omega}{2\hbar}}\left(q+i\frac{p}{m\omega}\right).
\end{equation*}
A basis of $\hilb{H}$ is given by the eigenvectors of the number
operator $\widehat{N}=\widehat{a}^\dagger\widehat{a}$ and will be
denoted by $\ket{n}$. 
The coherent states for the harmonic oscillator are defined as  
\begin{equation*}
  \ket{z}=\exp\left(-\frac{\betr{z^2}}{2t}\right)\sum_{n=0}^\infty
  \frac{z^n}{\sqrt{t^nn!}}\ket{n},\qquad z\in \C  
\end{equation*}
A system of two harmonic oscillators can be quantized on the
tensor product $\hilb{H}^{\text{kin}}\doteq \hilb{H}\otimes\hilb{H}$,
the annihilation operators of the respective oscillators are given by  
\begin{equation*}
  \widehat{a}_1\doteq \widehat{a}\otimes \one, \qqquad 
  \widehat{a}_2\doteq \one \otimes \widehat{a}, 
\end{equation*}
and similarly for the number operators
$\widehat{N}_1,\widehat{N}_2$. Analogously we have  
\begin{equation*}
  \ket{n_1,n_2}\doteq \ket{n_1}\otimes \ket{n_2},\qquad
  \ket{z_1,z_2}\doteq \ket{z_1}\otimes \ket{z_2},\qquad n_1,n_2\in
  \N_0,\, z_1,z_2 \in \C
\end{equation*}
and these vectors form dense subsets in $\hilb{H}^{\text{kin}}$. 

Let us now impose the constraint $C\doteq N_1-N_2$ forcing the
energies of the two oscillators to be equal. 
The kinematical phase space can be labeled by $(z_1,z_2)\in\C^2$, the
physical phase space by $z\in\C$, where the embedding of the latter in 
the former is given by 
\begin{equation}
\label{emb}
  \betr{z}=\betr{z_1}=\betr{z_2},\qquad z\betr{z}=z_1z_2.
\end{equation}
The quantization of the constraint is simply
\begin{equation}
\label{eqa5}
  \widehat{C}= \widehat{N}_1-\widehat{N}_2, 
\end{equation}
the physical subspace of $\hilb{H}^{\text{kin}}$ is given by  
\begin{equation*}
  \hilb{H}^{\text{phys}}=\overline{\spa \left\{\ket{n,n},\,n\in 
\N_0\right\}}.  
\end{equation*}
On this subspace, a new annihilation operator can be defined by
\begin{equation*}
  \widehat{a}_\otimes \doteq \widehat{a}_1
  \widehat{N}_1^{-\frac{1}{4}}\widehat{a}_2 \widehat{N}_2^{-\frac{1}{4}}
\end{equation*}
on $\spa \{\ket{n,n},\,n\in \N \}$ and 
$\widehat{a}_\otimes \ket{0,0}\doteq 0$.  
It fulfills
$\comm{\widehat{a}_\otimes}{\widehat{a}_\otimes^\dagger}=\one$.
Therfore we can define \textit{physical} coherent states as
\begin{equation*}
  \ket{z}_\otimes =\exp\left(-\frac{\betr{z^2}}{2}\right)\sum_{n=0}^\infty
  \frac{z^n}{n!}(\widehat{a}_\otimes^\dagger)^n\ket{0,0}.  
\end{equation*}
These are to be compared with the kinematical coherent states
$\ket{z_1,z_2}$, bearing in mind the identification \eqref{emb}. 

Let us consider the expectation values of the Dirac observables 
$\widehat{a}_1\widehat{a}_2$ and $\widehat{N}_1$
and more complicated ones constructed from them. We start with $N_1$: Clearly
\begin{equation*}
  \expec{\widehat{N}_1}_{\ket{z_1,z_2}}=\betr{z_1}^2=\betr{z}^2. 
\end{equation*}
On the other hand, one finds 
\begin{equation*}
  \expec{\widehat{N}_1}_{\ket{z}_\otimes}=\betr{z}^2, 
\end{equation*}
so in this case the expectation values agree \textit{exactly}. Now we
turn to $\widehat{a}_1\widehat{a}_2$: In the kinematical coherent states
\begin{equation}
\label{eqa3}
  \expec{\widehat{a}_1\widehat{a}_2}_{\ket{z_1,z_2}}=z_1z_2=z\betr{z}. 
\end{equation}
The expectation value in the physical coherent states is  
\begin{equation}
\label{eqa1}
  \expec{\widehat{a}_1\widehat{a}_2}_{\ket{z}_\otimes}= z
\exp\left(-\betr{z}^2\right)\sum_{n=0}^\infty \sqrt{n+1}
\frac{\betr{z}^{2n}}{n!}.
\end{equation}
The sum in this formula can not be determined in terms of elementary
functions. We can, however study its behavior for large $\betr{z}$. To this
end, let us define the function 
\begin{equation}
\label{eqa2}
F_\alpha(b)\doteq \sum_{n=1}^\infty
\frac{b^n}{n!}\left(\frac{n}{b}\right)^\alpha.   
\end{equation}
Then we can write \eqref{eqa1} as
\begin{equation*}
  \expec{\widehat{a}_1\widehat{a}_2}_{\ket{z}_\otimes}
  =z\betr{z}e^{-b}F_{\frac{3}{2}}(b) 
\end{equation*}
where $b=\betr{z}^2$.
To obtain an asymptotic formula for $F_\alpha(b)$ for large $b$,  
we approximate the factorial in \eqref{eqa2} by Stirlings 
formula and the discrete sum by an integral. We find  
\begin{equation*}
F_\alpha(b)\approx \sqrt{\frac{b}{2\pi}}\int_0^\infty
x^{\alpha-\frac{1}{2}} \exp \left(-bx(\ln x-1)\right)\,dx
\end{equation*}
The asymptotic behavior of this integral can be obtained by saddle
point methods (see for example \cite{asym}). We obtain 
\begin{equation}
\label{eqa4}
F_\alpha(b)=
e^b\left[1+\frac{1}{b}\left(\frac{1}{2}(\alpha-\frac{1}{2})^2
    -\frac{1}{8}\right)+O(b^{-2})\right]. 
\end{equation}
Therefore the expectation value \eqref{eqa1} is 
\begin{equation*}
  \expec{\widehat{a}_1\widehat{a}_2}_{\ket{z}_\otimes}
  =
  z\betr{z}\left(1+\frac{3}{8}\frac{1}{\betr{z}^2}+O(\betr{z}^{-4})\right).
\end{equation*}
Comparing this with \eqref{eqa3}, we see that the expectation values
in kinematical and physical coherent states disagree by a term of
order $1$. This is a small correction if $\betr{z}$ is large.   

Similar results can be obtained for more complicated functions of the Dirac
observables. Consider for example the operator $\widehat{a}_\otimes$. 
It can be written as
$\widehat{a}_1\widehat{a}_2\widehat{N}_1^{-1/4}\widehat{N}_2^{-1/4}$. 
In this case, the expectation value in the physical coherent states is 
trivial:
\begin{equation*}
  \expec{\widehat{a}_\otimes}_{\ket{z}_\otimes}=z.
\end{equation*}
On the other hand, we find
\begin{equation*}
  \expec{\widehat{a}_\otimes}_{\ket{z_1,z_2}}=\frac{z_1}{\sqrt{\betr{z_1}}}
\frac{z_2}{\sqrt{\betr{z_2}}}e^{-b_1}e^{-b_2}F_{\frac{3}{4}}(b_1)
F_{\frac{3}{4}}(b_2)
\end{equation*}
where $b_i=\betr{z_i}^2$. Using \eqref{eqa4} this can be simplified to 

\begin{equation*}
  \expec{\widehat{a}_\otimes}_{\ket{z_1,z_2}}= z
\left(1-\frac{3}{16}\frac{1}{\betr{z}^2}+O(\betr{z}^{-4})\right). 
\end{equation*}
Summarizing, we find that in the simple example of two harmonic
oscillators coupled by the constraint \eqref{eqa5}, expectation values
of Dirac observables in both, kinematical and physical coherent
states, can be computed to any desired order of accuracy. For some
observables, these expectation values agree. For others, there are
$\hbar$-corrections.
This indicates that kinematical coherent states always give the same 
answer to zeroth order as the dynamical ones and that the first 
corrections differ by a constant of proportionality of order unity so
that at least qualitatively 
we get a good idea of which corrections to expect in the exact
theory. 
\end{appendix}
\bibliographystyle{JHEP-2}
\bibliography{lqg2}        
\end{document}